\documentclass{article}

\usepackage{arxiv}
\usepackage{url}

\usepackage[american]{babel}

\usepackage{natbib} 
\usepackage{mathtools} 
\usepackage{booktabs} 
\usepackage{tikz} 
\usepackage{bm}
\usepackage{amssymb,amsmath,amsthm}

\newtheorem{theorem}{Theorem}

\usepackage{algorithm}
\usepackage{algorithmic}
\usepackage[switch]{lineno}
\usepackage{diagbox}

\usepackage{amssymb}
\usepackage{comment}
\usepackage{subfigure}
\usepackage{xcolor}
\definecolor{lightgray}{gray}{0.893}
\usepackage{colortbl}

\title{Fast Teammate Adaptation in the Presence of Sudden Policy Change}

\author{%
  Ziqian Zhang$^{1,}$\thanks{The first two authors contributed equally.}, Lei Yuan$^{1,2,*}$, Lihe Li$^1$, Ke Xue$^1$, Chengxing Jia$^{1, 2}$, Cong Guan$^1$, Chao Qian$^1$, {Yang Yu}$^{1,2,}$\thanks{Corresponding Author}\\
  $^1$ National Key Laboratory for Novel Software Technology, Nanjing University\\
  $^2$ Polixir.ai\\
  \texttt{\{zhangzq,yuanl,lilh,xuek,jiacx,guanc\}@lamda.nju.edu.cn}, \texttt{\{qianc,yuy\}@nju.edu.cn}
}

\begin{document}

\maketitle

\begin{abstract}
  In cooperative multi-agent reinforcement learning (MARL), where an agent coordinates with teammate(s) for a shared goal, it may sustain non-stationary caused by the policy change of teammates. Prior works mainly concentrate on the policy change during the training phase or teammates altering cross episodes, ignoring the fact that teammates may suffer from policy change suddenly within an episode, which might lead to miscoordination and poor performance as a result. We formulate the problem as an open Dec-POMDP, where we control some agents to coordinate with uncontrolled teammates, whose policies could be changed within one episode. Then we develop a new framework \textit{\textbf{Fas}t \textbf{t}eammates \textbf{a}da\textbf{p}tation (\textbf{Fastap})} to address the problem. Concretely, we first train versatile teammates' policies and assign them to different clusters via the Chinese Restaurant Process (CRP). Then, we train the controlled agent(s) to coordinate with the sampled uncontrolled teammates by capturing their identifications as context for fast adaptation. Finally, each agent applies its local information to anticipate the teammates' context for decision-making accordingly. This process proceeds alternately, leading to a robust policy that can adapt to any teammates during the decentralized execution phase. We show in multiple multi-agent benchmarks that Fastap can achieve superior performance than multiple baselines in stationary and non-stationary scenarios. 

\end{abstract}

\section{Introduction}\label{sec:intro}

Cooperative Multi-agent Reinforcement Learning (MARL) has shown great promise in recent years, where multiple agents coordinate to complete a specific task with a shared goal~\citep{oroojlooy2022review}, achieving great progress in various domains (e.g., path finding~\citep{sartoretti2019primal},  active voltage control~\citep{DBLP:conf/nips/WangXGSG21}, and dynamic algorithm configuration~\citep{xue2022multiagent}). Various methods emerge as promising solutions, including policy-based ones~\citep{maddpg,mappo}, value-based series~\citep{vdn,qmix}, and many variants like transformer~\citep{wen2022multiagent}, showing remarkable coordination ability in a wide range of tasks like StarCraft multi-agent challenge (SMAC), Google Research Football (GRF)~\citep{gorsane2022towards}, etc. 
Other works investigate different aspects, including  communication among agents~\citep{zhu2022survey}, model learning~\citep{wang2022model}, policy robustness~\citep{guo2022towards}, ad hoc teamwork~\citep{mirsky2022survey}, etc. 



However, one issue that can arise in MARL is non-stationarity~\citep{papoudakis2019dealing} caused by changes in teammates' policies.
Non-stationary is a hazardous issue for reinforcement learning, either in single-agent reinforcement learning (SARL)~\citep{Padakandla2019ReinforcementLA}, or MARL~~\citep{papoudakis2019dealing} settings, where the environment dynamic (e.g., transition or reward functions) of a learning system may change over time (inter- or intra-episodes).
Many solutions have been developed in SARL to relieve this problem, 
including meta-reinforcement learning~\citep{beck2023survey}, strategic retreat~\citep{DBLP:conf/case/DastiderL22}, sticky Hierarchical Dirichlet Process (HDP) prior~\citep{DBLP:conf/iclr/RenSJSWB22}, etc. 
The non-stationary in MARL is, however, much more complex, as we should consider the policy change caused by multiple teammates rather than the single environment dynamic change in SARL. 
The majority of works in MARL mainly focus on the non-stationary during the training phase~\citep{DBLP:journals/ai/AlbrechtS18,DBLP:conf/icml/Kim0RSAHLTH21}, the teammates' policy change across episodes ~\citep{qin2022multi,DBLP:conf/icml/HuLPF20},  or when perturbations happen~\citep{guo2022towards} (See related work in App.~A). However, the sudden policy change of teammates when deployed within an episode is never explored to the best of our knowledge, neither in problem formulation nor efficient algorithm design. Ignoring this issue would result in policy shift and even catastrophic miscoordination as agents' policies depend on other teammates in MARL~\cite{zhang2021multi}. On the other hand, the successful approaches used in SARL are unsuitable for the MARL setting because of the MARL's inherent characteristic (e.g., partial observability). This begs the question: Can we acquire a robust policy that can handle such changes and adapt to the new teammates' polices rapidly?
 


In this work, we aim to develop a robust coordination policy for the mentioned issue. 
Concretely, we formulate the problem as an Open Dec-POMDP, where we control multiple agents to coordinate with some uncontrolled teammates, whose policies could be altered unpredictably within one episode. Subsequently, we develop a new training framework Fastap, with which an agent can anticipate the teammates' identification via its local information. Specifically, as similar teammates might possess similarities in their identifications, learning a specific context for each teammate but ignoring the relationships among them could lead to trivial encodings. 
We thus assign them to different clusters via the Chinese Restaurant Process (CRP) to shrink the context search space. For the controlled coordinating policy training, we sample representative teammates to coordinate with by capturing their identifications into distinguishing contexts to augment the joint policy during the centralized training phase. Each agent then utilizes its local information to approximate the global context information. The mentioned processes proceed alternately, and we can finally obtain a robust policy to adapt to any teammates gradually during the decentralized execution phase.

For evaluation, we conduct experiments on different MARL benchmarks where the teammates' policy alter within one episode, including level-based foraging (LBF)~\citep{lbf}, Predator-prey (PP), Cooperative navigation (CN) from MPE~\citep{maddpg}, and a map created from StarCraft Multi-Agent Challenge (SMAC)~\citep{pymarl}. Experimental results show that the proposed Fastap can cluster teammates to distinguishing groups, learn meaningful context to capture teammates' identification, and achieve outstanding performance in stationary and non-stationary scenarios compared with multiple baselines.



\begin{figure*}
  \centering
  \includegraphics[width=0.99\textwidth]{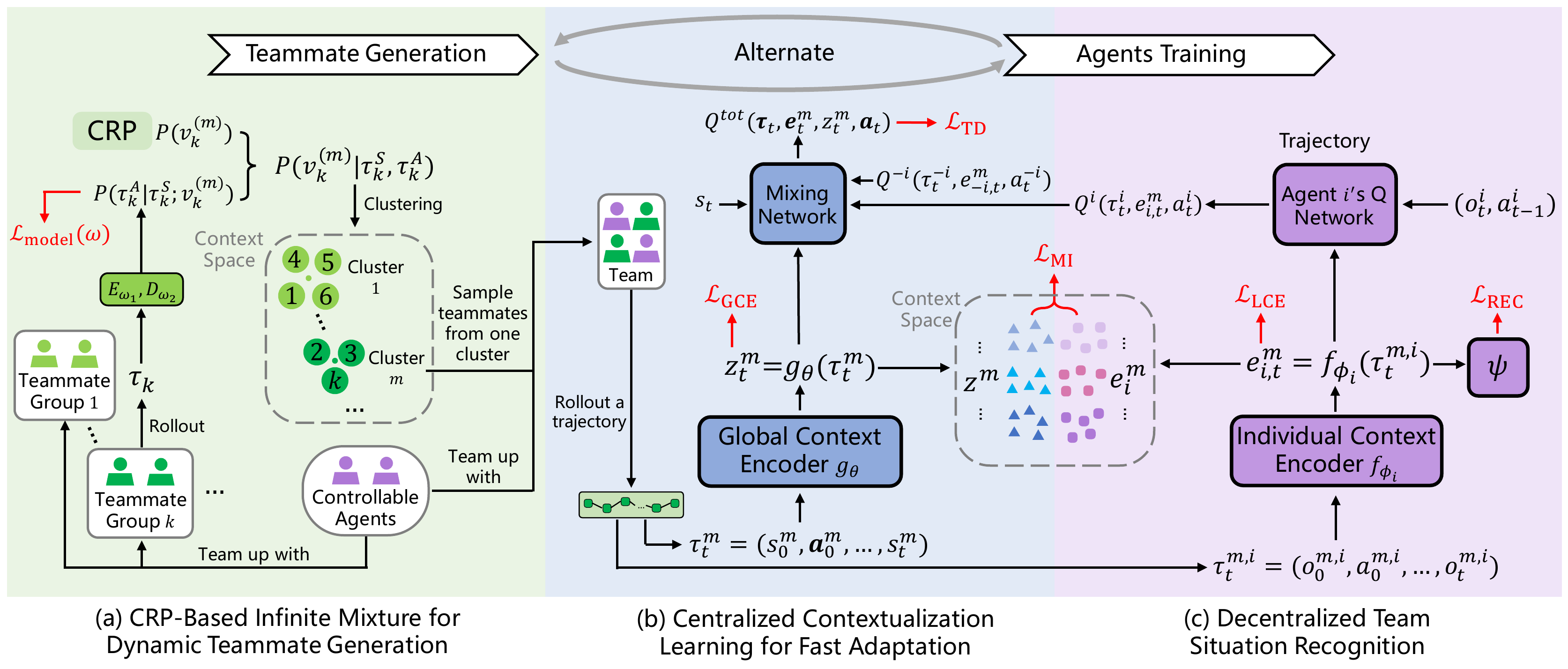}
  \caption{The overall framework of Fastap.}
  \label{Structure}
\end{figure*}
\section{Problem Formulation} 
The aim of this work is to train multiple controllable agents to interact with other teammates that might suddenly change their policies at any time step within one episode. Therefore we formalize the problem by extending the framework of Dec-POMDP~\citep{oliehoek2016concise} to an Open Dec-POMDP $\mathcal{M}=\langle \mathcal{N}, \mathcal{\bar N}, \mathcal{S}, \mathcal{A}, \mathcal{\bar A}, P, \Omega, O, R,\mathcal{U}, \gamma \rangle$. Here $\mathcal{N}=\{1,...,n\}$, $\mathcal{\bar N}=\{\bar{1},...,\bar{m}\}$ are the sets of controllable agents and uncontrollable teammates, respectively, 
$\mathcal{S}$ stands for the set of state, $\mathcal{A}=\mathcal{A}^1\times...\times\mathcal{A}^n$ and $\mathcal{\bar A}=\mathcal{A}^{\bar 1}\times...\times \mathcal{A}^{\bar m}$ are  the corresponding sets of joint actions for $\mathcal N$ and $\mathcal{\bar N}$, 
 $P$, $O$, $R$ denote the corresponding transition, observation, and reward functions,  $\Omega$ is the set of  observations, $\gamma \in [0, 1)$ is the discounted factor, and $\mathcal{U}$ is a probability distribution  used to control the frequency of sudden change.


At the beginning of each episode, the set of uncontrollable teammates that participate in the cooperation at the very start is denoted by $\mathcal{\bar N}_0\in \mathcal{P}(\mathcal{\bar N})$, where $\mathcal{P}(\cdot)$ stands for the power set, and the waiting time is represented by $u_0\sim \mathcal{U}$.
At each time step $t$, $u_t=u_{t-1}-1$ and $\mathcal{\bar N}_t=\mathcal{\bar N}_{t-1}$ are updated. If $u_t\leq 0$, it will be resampled from $\mathcal{U}$, and a brand new set of uncontrollable teammates $\mathcal{\bar N}_t\in \mathcal{P}(\mathcal{\bar N})$ will replace the previous one. Meanwhile,
controllable agent $i$ receives the observation $o^i=O(s, i)$ and outputs action $a^i\in \mathcal{A}^i$, and so do the uncontrollable teammates. Notice that the number of uncontrollable teammates is changeable in one episode. The joint action $(\boldsymbol{a}, \boldsymbol{\bar a})$ leads to the next state $s'\sim P(\cdot|s, (\boldsymbol{a}, \boldsymbol{\bar a}))$ and a shared reward $R(s, (\boldsymbol{a}, \boldsymbol{\bar a}))$, where $\boldsymbol{a}=(a^1, ..., a^n)\in \mathcal{A}$ and $\boldsymbol{\bar {a}}\in \{(a^{\bar i})_{\bar i\in {\bar N}}|a^{\bar i}\in \mathcal{A}^{\bar i}, \bar N\in \mathcal{P}(\mathcal{\bar N})\}$.  To relieve the partial observability, the trajectory history $(o^i_1, a^i_1,...o_{t-1}^i, a_{t-1}^i, o_t^i)$ of agent $i$ until time step $t$  is encoded into  $\tau^i_t$ by GRU~\citep{gru}. Under an Open Dec-POMDP, we aim to find an optimal policy when uncontrollable teammates suffer from a sudden change. Then, with $\boldsymbol{\tau}_t=\langle \tau^1_t,...,\tau^n_t\rangle$, the formal objective is to find a joint policy $\boldsymbol{\pi}(\boldsymbol{\tau}_t, \boldsymbol{a})$ for controllable agents, which maximizes the global value function $Q_{\text{tot}}^{\boldsymbol{\pi}}(\boldsymbol{\tau}, \boldsymbol{a})=\mathbb{E}_{s, \boldsymbol{a}, \boldsymbol{\bar a}}[\sum_{t=0}^\infty \gamma^t R(s, (\boldsymbol{a}, \boldsymbol{\bar a}))|s_0=s, \boldsymbol{a}_0=\boldsymbol{a}, \boldsymbol{\pi}, \boldsymbol{\bar \pi}]$, where $\boldsymbol{\bar \pi}$ is the unknown joint policy of uncontrollable teammates.

\section{Method} \label{sec:3 method}
In this section, we will present the detailed design of Fastap (see Fig.~\ref{Structure}), a novel multi-agent policy learning approach that enables controllable agents to handle the sudden change of teammates' polices and adapt to new teammates rapidly. First, we design an infinite mixture model that formulates the distribution of continually increasing teammate clusters based on the Chinese Restaurant Process (CRP)~\citep{crp} (Sec~\ref{sec:3.1 crp mixture} and Fig.~\ref{Structure}(a)). Next, we introduce the centralized context encoder learning objective for fast adaption (Sec~\ref{sec3.2: context} and Fig.~\ref{Structure}(b)). Finally, considering the popular CTDE paradigm in cooperative MARL, we train each controllable agent to recognize and adapt to the teammate situation rapidly according to its local information (Sec~\ref{sec3.3: opt} and Fig.~\ref{Structure}(c)).  
\subsection{CRP-based Infinite Mixture for Dynamic Teammate Generation} \label{sec:3.1 crp mixture}
To adapt to the sudden change in teammates with diverse behaviors in one episode rapidly during evaluation, we expect to maintain a set of diverse policies to simulate the possibly encountered teammates in the training phase. Nevertheless, it is unreasonable and inefficient to consider every newly generated group of teammates as a novel type while ignoring the similarities among them. This approach lacks scalability in a learning process where teammates are generated incrementally, and it may lead to reduced training effectiveness if teammates with similar behavior are generated.
Accordingly, we expect to acquire clearly distinguishable boundaries of teammates' behaviors by applying a behavior-detecting module to assign teammate groups with similar behaviors to the same cluster. 
To tackle the issue, an infinite Dirichlet Process Mixture (DPM) model~\citep{lee2020neural} could be applied due to its scalability and flexibility in the number of clusters. 
Concretely, we can formulate the teammate generation process as a stream of teammate groups with different trajectory batch $\mathcal{D}_1, \mathcal{D}_2, ...$ where each batch $\mathcal{D}_k$ is a set of trajectories $\tau = (s_0, \boldsymbol{a}_0 ..., s_T)$ sampled from the interactions between the $k^{\text{th}}$ teammate group and the environment, and $T$ is the horizon length. Considering the difficulty of trajectory representation due to its high dimension, we utilize a trajectory encoder $E_{\omega_1}$ parameterized by $\omega_1$ to encode $\tau$ into a latent space. Specifically, we partition the trajectory $\tau$ into $\tau^S=(s_0, ... s_{T-1}, s_T)$ and $\tau^A=(\boldsymbol{a}_0, ..., \boldsymbol{a}_{T-1})$, 
and a transformer architecture is applied to extract features from the trajectory and represent it as $v=E_{\omega_1}(\tau)$. For the $k^{\text{th}}$ teammate group generated so far, $v_k = \mathbb{E}_{\tau_k\sim \mathcal{D}_k}[E_{\omega_1}(\tau_k)]$ will be used to represent its behavioral type, and $\bar v^m$ is the mean value of the $m^{\text{th}}$ cluster.


If $M$ clusters are instantiated so far, the cluster that the $k^{\text{th}}$ teammate group belongs to will be inferred from the assignment $P(v_k^{(m)}|\tau_k)=P(v_k^{(m)}|\tau^S_k, \tau^A_k), m=1,..., M, M+1$, where $v_k^{(m)}$ denotes that the $k^{\text{th}}$ group belongs to the $m^{\text{th}}$ cluster based on its representation $v_k$. 
The posterior distribution can be written as:
\begin{equation}
    \begin{aligned}
        P(v_k^{(m)}|\tau^S_k, \tau^A_k)\propto P(v_k^{(m)})P(\tau_k^A|\tau_k^S; v_k^{(m)}),
    \end{aligned}
\end{equation}

we apply CRP~\citep{crp} to instantiate the DPM model as the prior.  
Specifically, for a sequence of teammate groups whose representations are $[v_1, v_2, ... v_k, ...]$, the prior $P(v_k^{(m)})$ is set to be:
\begin{equation}
    \begin{aligned}
        P(v_k^{(m)}) = \begin{cases}
        \frac{n^{(m)}}{k-1+\alpha}, \quad m\leq M\\
        \frac{\alpha}{k-1+\alpha},\quad m=M+1,
        \end{cases}
    \end{aligned}
    \label{priori}
\end{equation}
where $n^{(m)}$ denotes the number of teammate groups belonging to the $m^{\text{th}}$ cluster, $M$ is the number of clusters instantiated so far, $\sum_{m=1}^M n^{(m)}=k-1$, and $\alpha>0$ is a concentration hyperparameter that controls the probability of the instantiation of a new cluster. 

To estimate the predictive likelihood $P(\tau_k^A|\tau_k^S; v_k^{(m)})$, we use an RNN-based decoder $D_{\omega_2}$ that takes $\tau_k^S, v_k^{(m)}$ as input and predicts $\tau_k^A$. The decoder represents each sample as an Gaussian distribution $\mathcal{N}(\mu(\tau_{t}^S, v), \sigma^2(\tau_{t}^S, v))$ where $\tau_t^S=(s_0, ..., s_t)$, such that
\begin{equation}
    \begin{aligned}
        P(\tau_k^A|\tau_k^S; v_k^{(m)})=& D_{\omega_2}(\tau_k^A|\tau_k^S; v_k^{(m)})\\
        =& \prod_{t=1}^T D_{\omega_2}(\boldsymbol{a}_t^k|\tau_{k, t}^S, v_k^{(m)}),\\
        \text{where}~ v_k^{(m)}=&\begin{cases}
            \frac{n^{(m)}\bar{v}^{m}+v_k}{n^{(m)}+1}\quad m\leq M\\
            v_k\quad\quad\quad\quad\,\, m=M+1.
        \end{cases}
    \end{aligned}
    \label{likelihood}
\end{equation}

Combing the estimated prior Eqn.~(\ref{priori}) and predictive likelihood Eqn.~(\ref{likelihood}), we are able to decide which cluster the $k^{\text{th}}$ teammate group belongs to and thus acquire clearly distinguishable boundaries of teammates' behavior. After the assignment, the mean value of the $m^{\text{th}}$ cluster will also be updated. Meanwhile, to force the learned representation $v$ to capture the behavioral information of each teammate group and estimate the predictive likelihood more precisely, the encoder $E_{\omega_1} $and decoder $D_{\omega_2}$ are optimized as:
\begin{equation}
    \begin{aligned}
        \mathcal{L}_{\text{model}}(\boldsymbol{\omega}) = -\log\mathbb{E}_{\tau\sim\cup_{k=1}^K\mathcal{D}_k}[D_{\omega_2}(\tau^A|\tau^S; E_{\omega_1}(\tau))],
    \end{aligned}
\end{equation}
where $K$ is the number of teammate groups generated so far, $\boldsymbol{\omega}=(\omega_1, \omega_2)$. The encoder and decoder are optimized while generating teammate groups (see details in App.~B.1).

\subsection{Centralized Contextualization Learning for Fast Adaptation}\label{sec3.2: context}
After gaining the generated teammates divided into different clusters, this part aims to train a robust policy to handle sudden teammate change and rapidly adapt to the new teammates via conditioning the controllable agents' policies on other teammates' behavior. Despite the diversity and complexity that unknown teammates' behavior exhibits, the CRP formalized before helps acquire clearly distinguishable boundaries based on teammates' behavioral types with regard to high-level semantics. 
Inspired by Environment Sensitive Contextual Policy Learning (ESCP)~\citep{escp}, which aims to guide the context encoder to identify and track the sudden change of the environment rapidly, we expect to utilize a global context encoder $g_\theta$ and local context encoder $\{f_{\phi_i}\}_{i=1}^n$ to embed the historical interactions into a compact but informative representation space. The encoders are supposed to identify a new type of teammate fast so as to recognize the sudden change in time, and we can optimize the encoder by proposing an objective that helps the encoder's output coverage to the oracle rapidly at an early time and keep consistent for the remaining steps.

During centralized training phase, we set $z_t^m=g_\theta(\tau_t^m)$, where $\tau_t^m=(s_0^m, \boldsymbol{a}_0^m, ..., s_t^m)$ is generated based on the interactions between the paired joint policy $(\boldsymbol{\pi}, \boldsymbol{\bar \pi}^m)$ and the environment, 
and $\boldsymbol{\bar \pi}^m$ is the joint policy of uncontrollable teammates belonging to the $m^{\text{th}}$ cluster. Notice that the cluster of teammates is chosen at the beginning of each episode and will not change during training, and sudden change of teammates only happens during evaluation. We can acquire the empirical optimization objective of  $g_\theta$ as:
\begin{equation}
    \begin{aligned}
        \mathcal{L}_{\text{GCE}} = \sum_{m=1}^M\mathbb{E}[||z^m_t-\bar z^m||_2^2]-\log\det(R_{\{\bar z^m\}}),
    \end{aligned}
    \label{loss_ce}
\end{equation}
where $\bar z^m$ is the moving average of all past context vectors used for stabilizing the training process, $\theta$ is the parameter of the global context encoder $g_\theta$, $\det(\cdot)$ denotes the matrix determinant, and $R_{\{\bar z^m\}}$ is a relational matrix. Intuitively, the objective expects to help the encoder's output coverage rapidly at an early time and keep it consistent for the remaining steps. Specifically, the former part forces $z_t^m$ to converge fast and stably in one episode, and the latter pushes the expectation of $z_t^m$ to a set of separable but representative latent vectors. The full derivation can be found in App.~B.2.

In practice, a recurrent neural network is applied to instantiate $g_{\theta}$, which takes $\tau_t^m=(s_0^m, \boldsymbol{a}^m_0, ..., s_t^m)$ as input and outputs a multivariate Gaussian
distribution $\mathcal{N}(\mu_\theta(\tau_t^m), \sigma^2_\theta(\tau_t^m))$. Thus the teammates context is obtained from the Gaussian distribution with the reparameterization trick by $z_t^m\sim g_\theta(\tau_t^m)$. As we can apply Fastap to any value-based methods, the global embedding $z_t^m$ could also be integrated into the centralized network. Similarly, the local embedding $e_t^{m, i}$ and local trajectory $\tau_t^{m, i} $ will also be concatenated to calculate the local Q-value $Q^i(\tau_t^{m, i}, e_t^{m, i}, \cdot)$, where the optimization of the local context encoder will be explained in detail in the next part. Therefore, the TD loss $\mathcal{L}_{\text{TD}}=[r_t^m+\gamma\max_{\boldsymbol{a}_{t+1}^m} \bar Q_{\text{tot}}(s_{t+1}^m, \boldsymbol{e}_{t+1}^m, z_{t+1}^m, \boldsymbol{a}_{t+1}^m)- Q_{\text{tot}}(s_{t}^m, \boldsymbol{e}_{t}^m, z_{t}^m, \boldsymbol{a}_{t}^m)]$ is utilized to accelerate the centralized contextualization learning, where $\bar Q_{\text{tot}}$ is periodically updated target Q network, and $\boldsymbol{e}_{t}^m=(\boldsymbol{e}_t^{m,i})_{i=1}^n$. The overall optimization objective of $g_\theta$ can thus be derived:
\begin{equation}
    \begin{aligned}
        \mathcal{L}_{\text{ADAP}} = \mathcal{L}_{\text{TD}} +\alpha_{\text{GCE}}\mathcal{L}_{\text{GCE}},
    \end{aligned}
    \label{loss_gce}
\end{equation}
where $\alpha_{\text{GCE}}$ is an adjustable hyper-parameter to balance the two optimization objective.

 \begin{figure*}[!ht]
  \centering
  \includegraphics[width=0.99\textwidth]{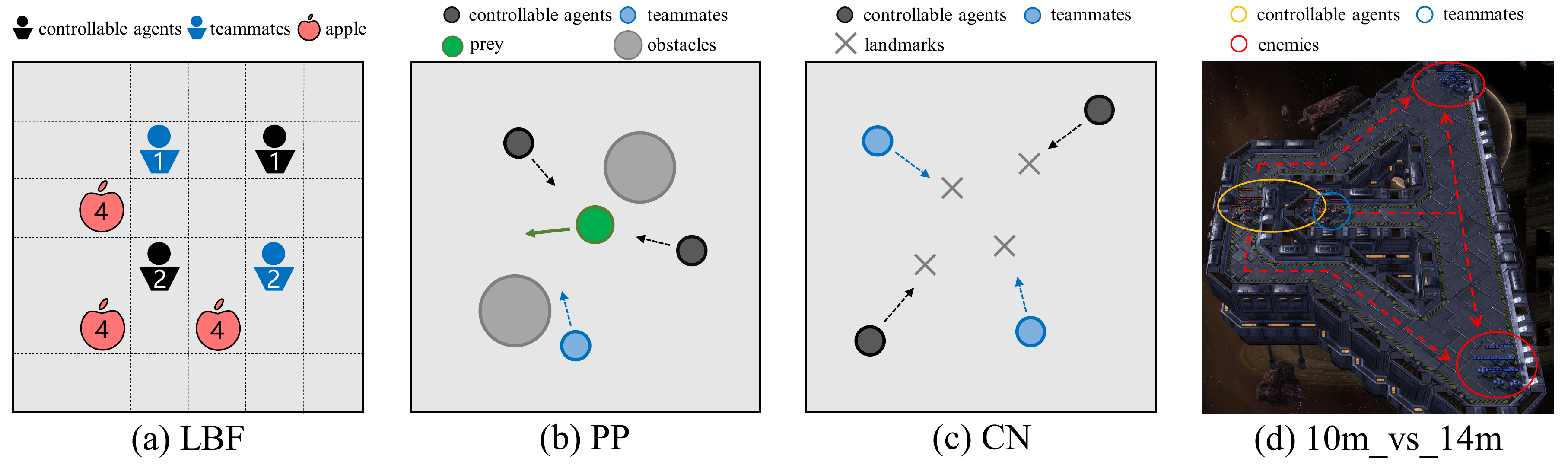}
  \caption{Experimental environments used in this paper.}
  \label{envs}
\end{figure*}

\subsection{Decentralized Team Situation Recognition and Optimization}\label{sec3.3: opt}
Despite the fact that optimizing Eqn.~(\ref{loss_gce}) helps obtain compact and representative representations $z_t^m$ that could guide individual policies to adapt to teammate sudden change rapidly, partial observability of MARL will not allow agents that execute in a decentralized manner to obtain $z_t^m$ encoded from the global state-action trajectory. Thus, we equip each agent $i$ with a local encoder $f_{\phi_i}$ to recognize the team situation. Concretely, the network architecture of $f_{\phi_i}$ is similar to $g_{\theta}$, $f_{\phi_i}$ takes local trajectory $\tau^{m, i}_t=(o_0^{m, i}, a_0^{m, i},..., o_t^{m, i})$ as input and outputs $e_{t}^{m, i}\sim\mathcal{N}(\mu_{\phi_i}(\tau_t^{m,i}), \sigma^2_{\phi_i}(\tau_t^{m,i}))$. To make $e_{t}^{m, i}$ informatively consistent with $z_t^m$, we introduce a mutual information (MI) objective by maximizing the MI $\mathcal{I}(e_{t}^{m, i};z_t^m
|\tau_t^{m,i})$ between $e_{t}^{m, i}$ and $z_t^m$ conditioned on the agent $i$'s local trajectory $\tau^{m, i}_t$. Due to the difficulty and feasibility of estimating the conditional distribution directly, variational distribution $q_{\xi}(e_{t}^{m, i}|z_t^m, \tau^{m, i}_t)$ is used to approximate the conditional distribution $p(e_{t}^{m, i}|z_t^m, \tau^{m, i}_t)$. Inspired by the information bottleneck~\citep{DBLP:conf/iclr/AlemiFD017}, we would derive a tractable lower bound of MI objective:
\begin{equation}
    \begin{aligned}
        &\mathcal{I}(e_{t}^{m, i};z_t^m
|\tau_t^{m,i})\geq \\
&\quad \mathbb{E}_{\mathcal{D}}[\log q_{\xi}(e_{t}^{m, i}|z_t^m, \tau^{m, i}_t)]+\mathcal{H}(e_{t}^{m, i}|\tau^{m, i}_t),
    \end{aligned}
\end{equation}
where $\mathcal{H}(\cdot)$ denotes the entropy, and variables of the distributions are sampled from the experience replay buffer $\mathcal{D}$. We defer the full derivation to App.~B.3. We can now rewrite the MI objective as:
\begin{equation}
    \begin{aligned}
        &\mathcal{L}_{\text{MI}}=\\
        &\quad \sum_{m=1}^M\sum_{i=1}^n  \mathbb{E}_{\mathcal{D}}[\log q_{\xi}(e_{t}^{m, i}|z_t^m, \tau^{m, i}_t)]+\mathcal{H}(e_{t}^{m, i}|\tau^{m, i}_t),
    \end{aligned}
\end{equation}

the mentioned symbols are defined similarly as Eqn.~(\ref{loss_ce}). To facilitate the learning process, two local auxiliary optimization objectives are further designed. On the one hand, we expect $e_{t}^{m, i}$ to recognize the team situation and adapt to new teammates that change suddenly as $z_t^m$ does:
\begin{equation}
    \begin{aligned}
        \mathcal{L}_{\text{LCE}} = \sum_{m=1}^M\sum_{n=1}^n\mathbb{E}[||e^{m,i}_t-\bar e^{m, i}||_2^2]-\log\det(R_{\{\bar e^{m, i}\}}).
    \end{aligned}
\end{equation}
On the other hand, to derive the descriptive representation $e^{m, i}_t$ of the specific team situation,  we hope $e^{m, i}_t$ can learn the relationship between controllable agents and the teammates. Therefore, we expect $e^{m,i}_t$ to reconstruct the observations and actions taken by teammates:
\begin{equation}
    \begin{aligned}
        \mathcal{L}_{\text{REC}} = \sum_{m=1}^M\sum_{n=1}^n\mathbb{E}_{\mathcal{D}}[-\log h_{\psi_i}(\boldsymbol{\bar o}_t^m, \boldsymbol{\bar a}_t^m|e_t^{m, i})],
    \end{aligned}
\end{equation}
where $h$ is parameterized by $\psi_i$ for each agent $i$.
As $e_t^{m, i}$ and $\tau_t^{m, i}$ will be concatenated into the input of individual Q network $Q^i(\tau_t^{m, i}, e_t^{m, i},\cdot)$, the TD loss $\mathcal{L}_{\text{TD}}$ is also utilized to 
promote the learning of local context encoder. Thus, the optimization objective  becomes:
\begin{equation}
    \begin{aligned}
        \mathcal{L}_{\text{DEC}} = \mathcal{L}_{\text{TD}}+\alpha_{\text{MI}}\mathcal{L}_{\text{MI}}+\alpha_{\text{LCE}}\mathcal{L}_{\text{LCE}}+\alpha_{\text{REC}}\mathcal{L}_{\text{REC}},
    \end{aligned}
\end{equation}
where $\alpha_{\text{MI}}, \alpha_{\text{LCE}}, \alpha_{\text{REC}}$ are the corresponding adjustable hyperparameters of the three objectives. 
\section{Experiments}
In this section, we design extensive experiments for the following questions:  1) Can Fastap achieve high adaptability and generalization ability when encountering teammate sudden change compared to other baselines in different scenarios, and how each component influences its performance (Sec.~\ref{results}) ? 2) Can CRP help 
acquire distinguishable boundaries of teammates' behaviors, and what team situation representation is learned by Fastap (Sec.~\ref{analysis})? 3) What transfer ability Fastap reveals, and how does each hyperparameter influence its coordination capability (Sec.~\ref{bonus})?

\begin{figure*}
\setlength{\abovecaptionskip}{0cm}
\centering
    \subfigure[LBF (Stationary)]{
    \label{lbf_static}
      \includegraphics[height=28.5mm]
      {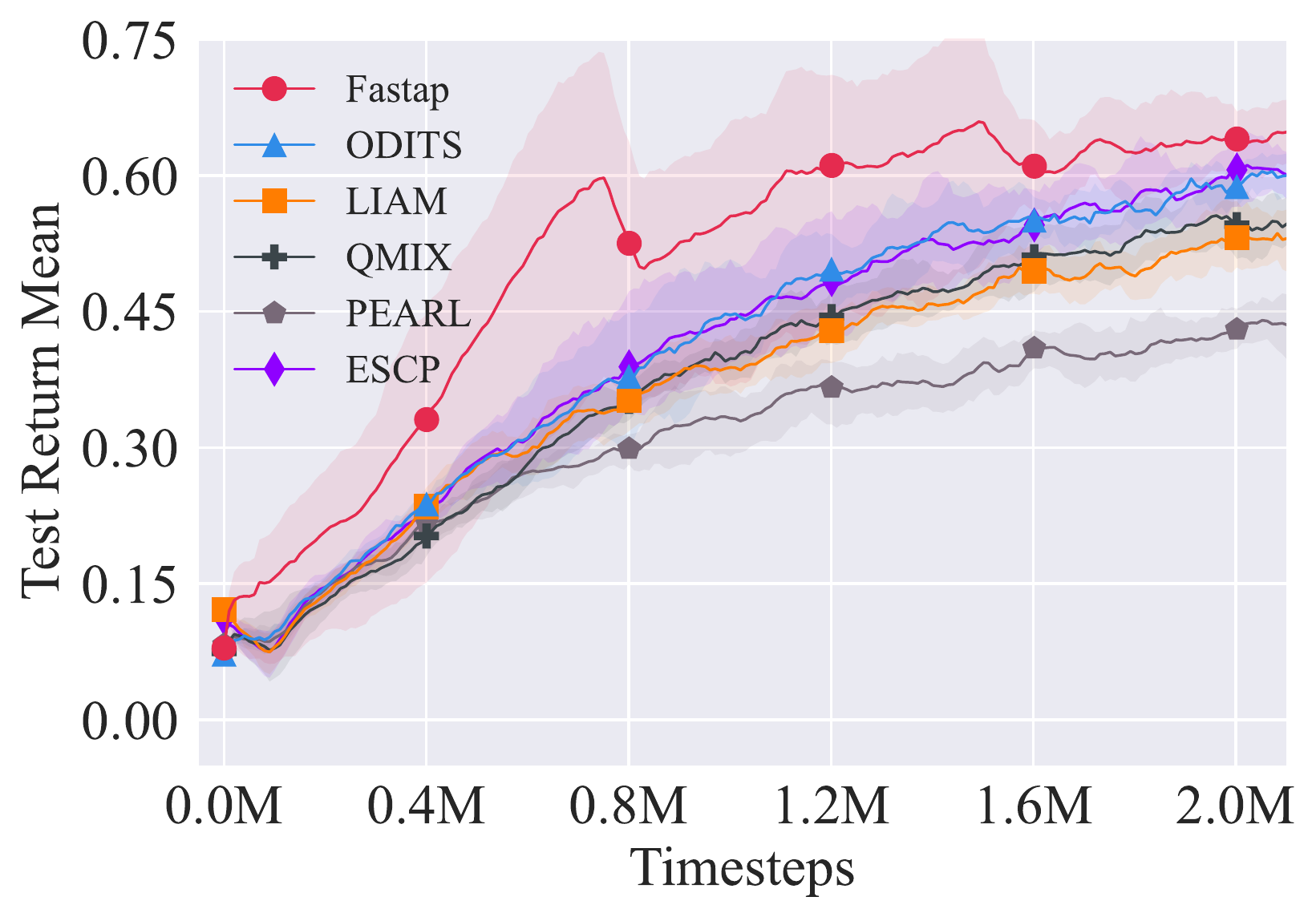}
    }
    \subfigure[PP (Stationary)]{
    \label{simple_tag_static}
      \includegraphics[height=28.5mm]
      {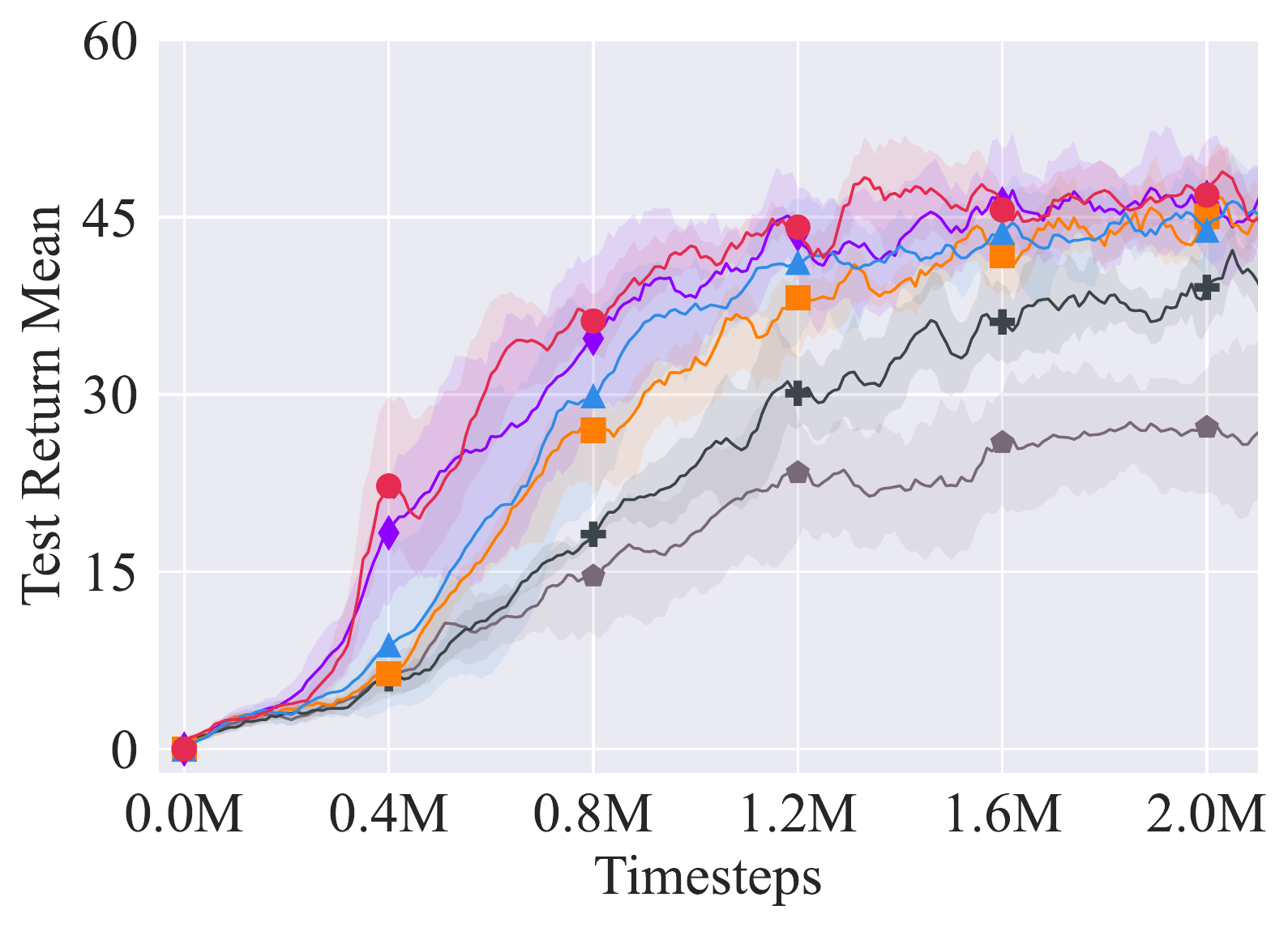}
    }
    \subfigure[CN (Stationary)]{
    \label{simple_spread_static}
      \includegraphics[height=28.5mm]
      {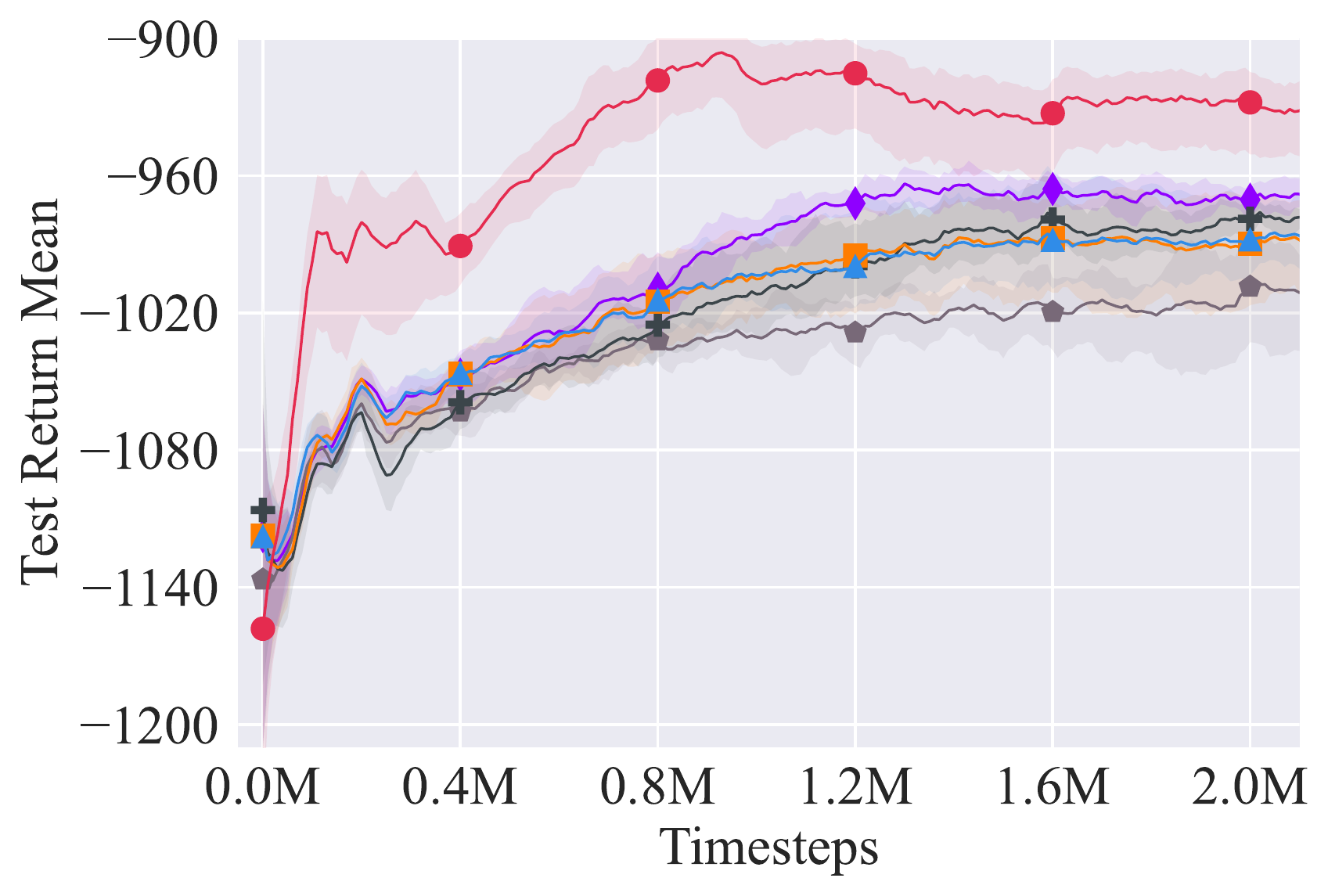}
    }
    \subfigure[10m\_vs\_14m (Stationary)]{
    \label{smac_static}
      \includegraphics[height=28.5mm]
      {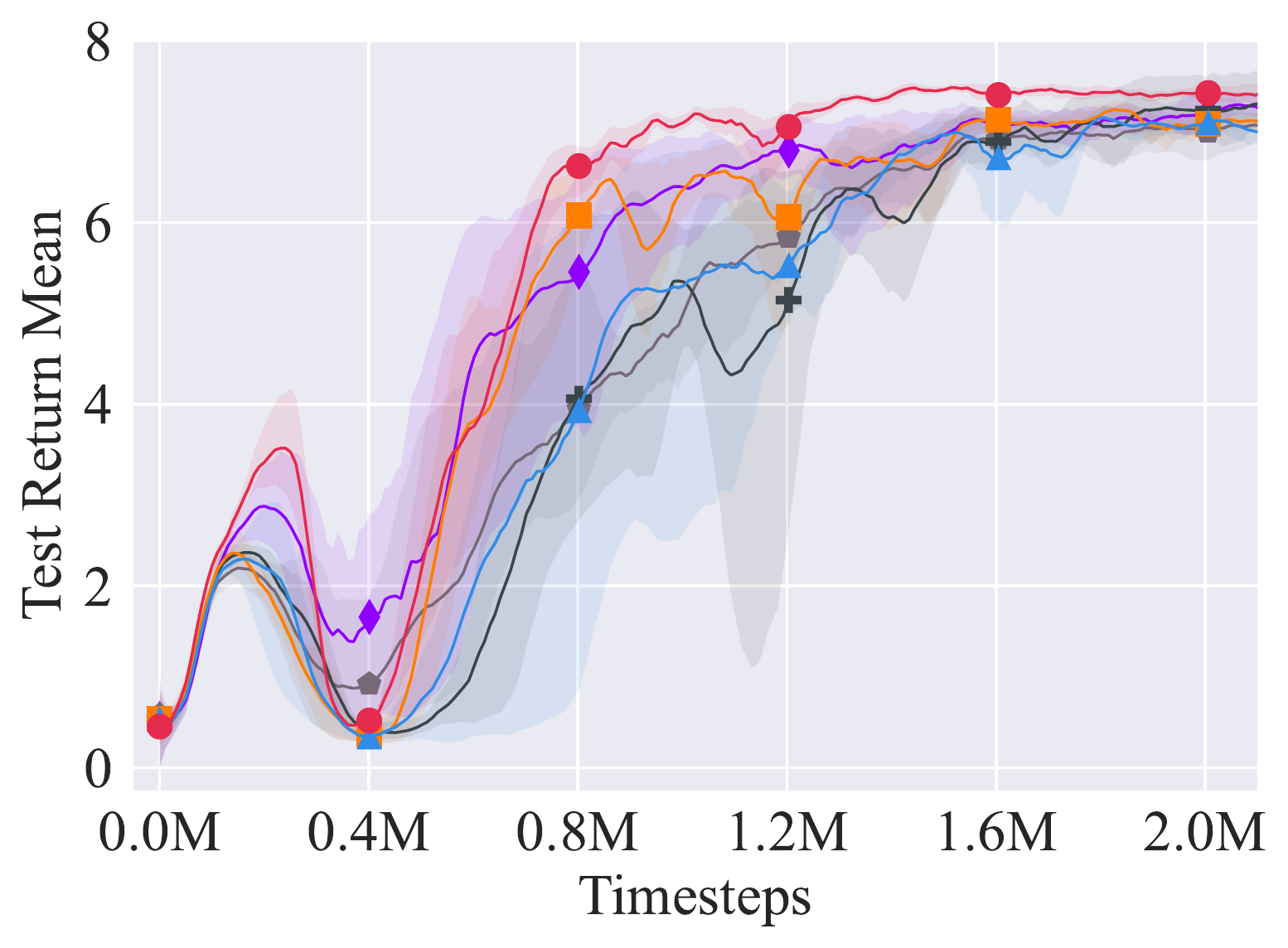}
    }

    \subfigure[LBF (Non-Sta.)]{
    \label{lbf_sudden}
      \includegraphics[height=28.5mm]
      {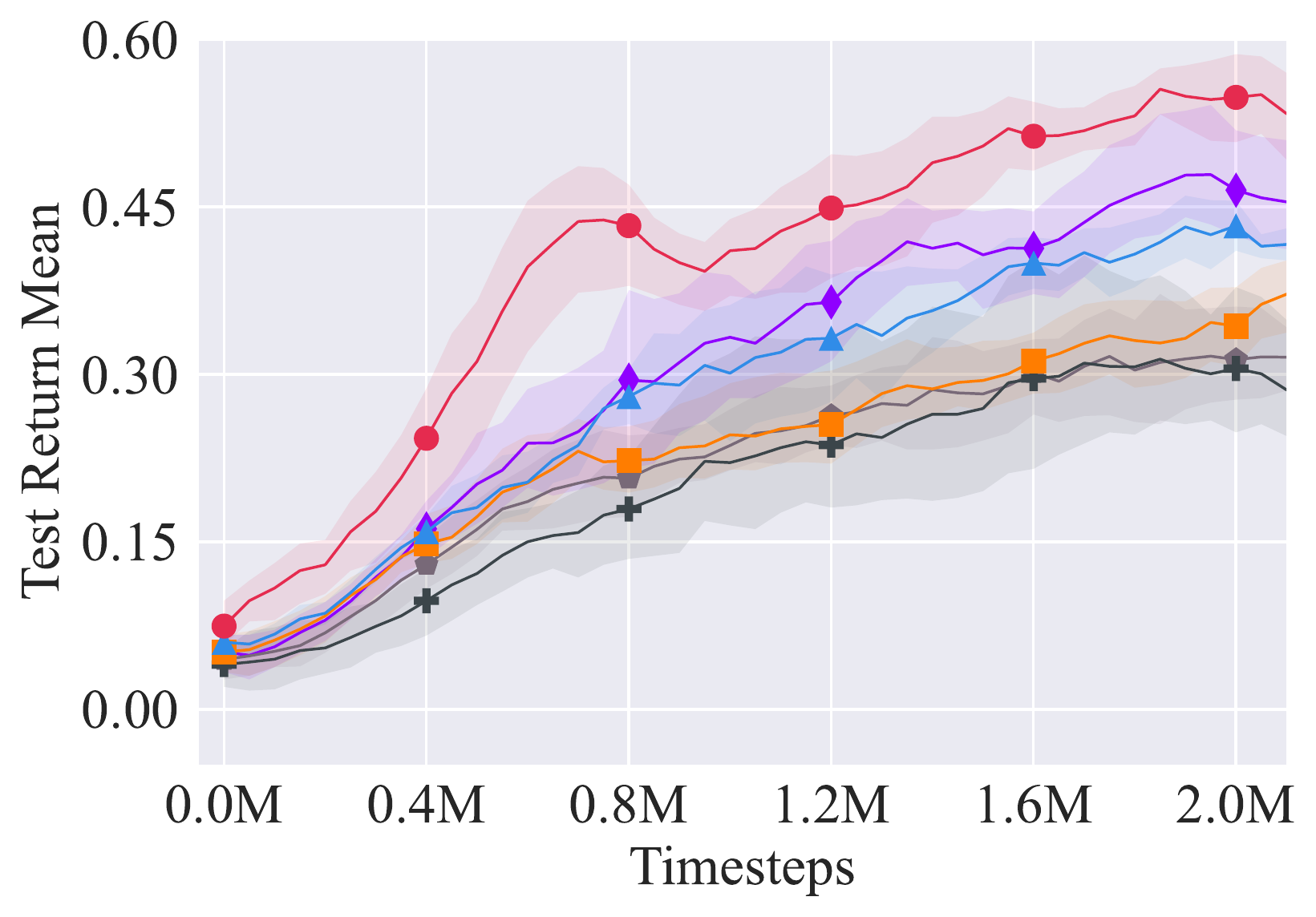}
    }
    \subfigure[PP (Non-Sta.)]{
    \label{simple_tag_sudden}
      \includegraphics[height=28.5mm]
      {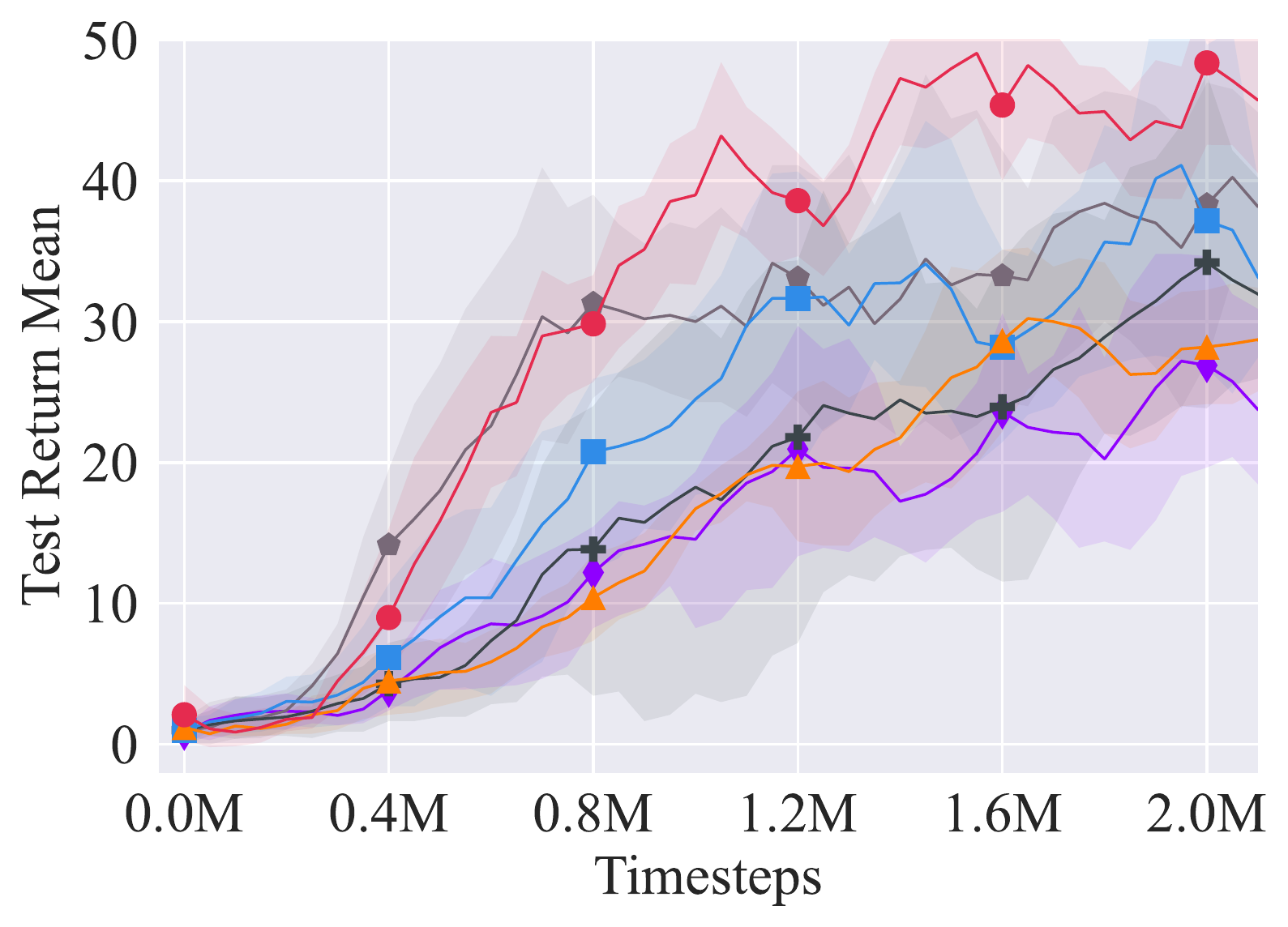}
    }
    \subfigure[CN (Non-Sta.)]{
    \label{simple_spread_sudden}
      \includegraphics[height=28.5mm]
      {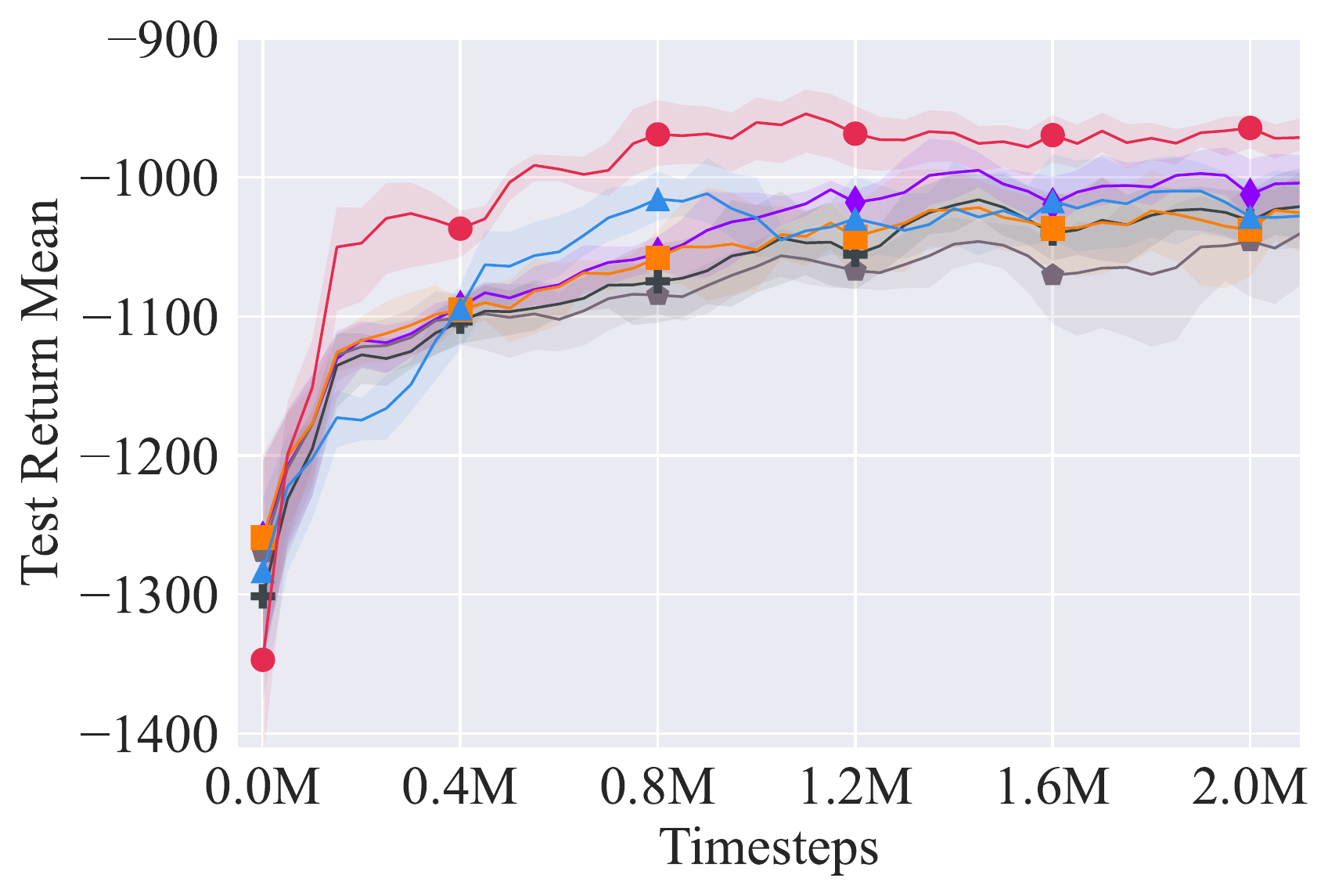}
    }
    \subfigure[10m\_vs\_14m (Non-Sta.)]{
    \label{smac_sudden}
      \includegraphics[height=28.5mm]
      {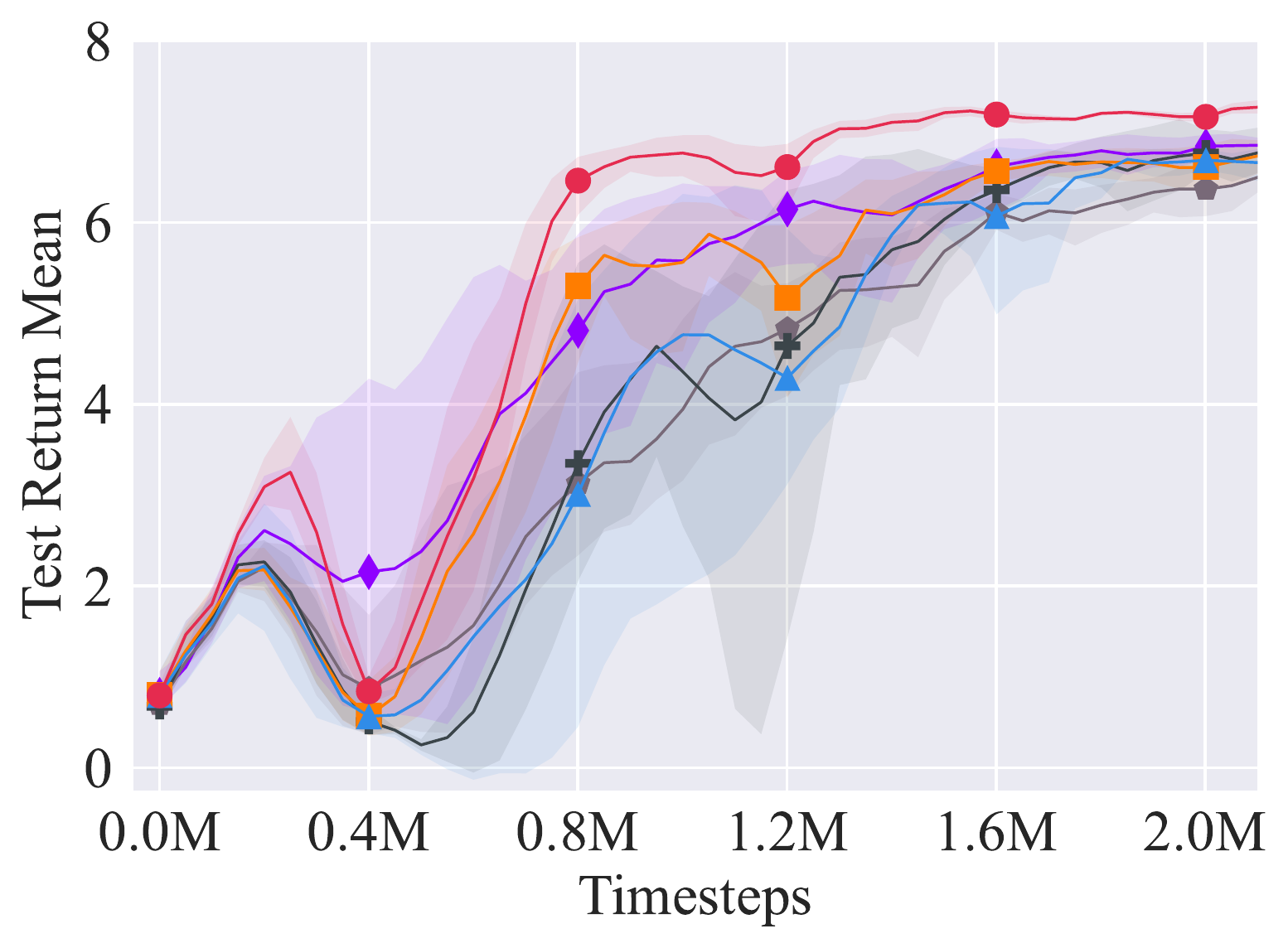}
    }  
  \vspace{.5em}
    \caption{Performance comparison with baselines on multiple benchmarks.}
  \label{main_exp}
\end{figure*}

\subsection{Environments and Baselines} \label{envandbas}
We select four multi-agent tasks as our environments, as shown in Fig.~\ref{envs}. Level Based Foraging (LBF)~\citep{lbf} is a cooperative grid world game with agents that are rewarded if they concurrently navigate to the food and collect it.  Predator-prey (PP) and Cooperative navigation (CN) are two scenarios coming from the MPE environment~\citep{maddpg}, where multiple agents (predators) need to chase and encounter the adversary agent (prey) to win the game in PP, and in CN, multiple agents are trained to move towards landmarks while avoiding collisions with each other.
We also create a map 10m\_vs\_14m from SMAC~\citep{pymarl}, where 10 allies are spawned at different points to attack  14 enemies to win.


For baselines, we consider multiple ones and implement them to a popular valued-based method QMIX~\citep{qmix} for comparisons, including (1) the vanilla QMIX without any extra design;
(2) Meta-learning SARL methods: PEARL~\citep{PEARL} uses recently collected context to infer a probabilistic variable describing the task; ESCP~\citep{escp} copes with the sudden change in the environment by learning a context-sensitive policy; 
(3) Context-based MARL approaches: LIAM~\citep{LIAM} predicts teammates' current behaviors based on local observation history to relieve non-stationary in the training phase; ODITS~\citep{ODITS} applies a centralized ``teamwork situation encoder'' for end-to-end learning to adapt to arbitrary teammates across episodes.  More details about the environments and baselines, and Fastap are illustrated in App.~C, and App.~D, respectively.


\subsection{Competitive Results and Ablations} \label{results}

\begin{figure}
  \centering
\includegraphics[width=0.48\textwidth]{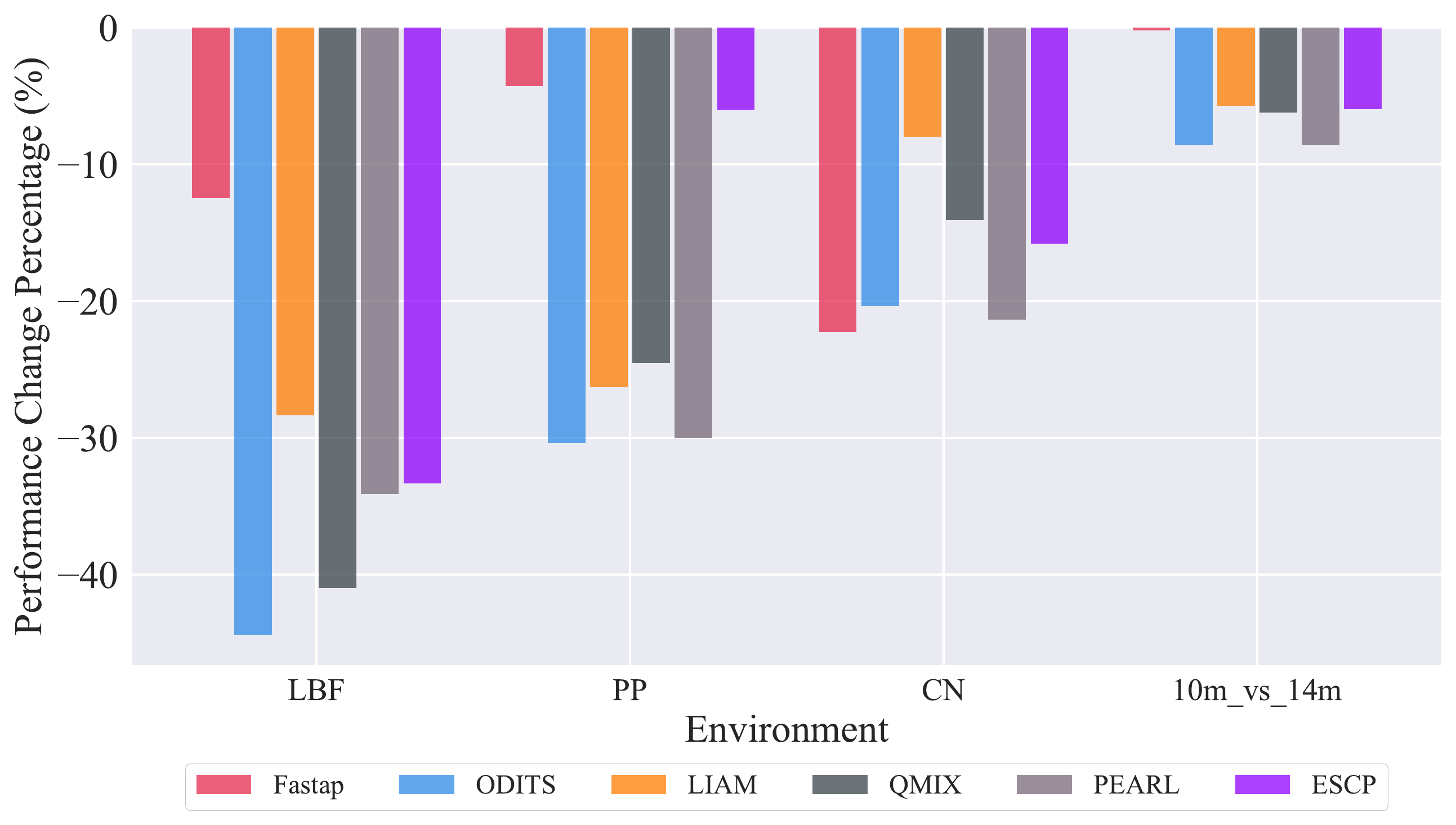}
  \caption{Performance difference in stationary and non-stationary conditions.
  The value is the difference in the performance under non-stationary and stationary settings w.r.t. the best return.
  }
  \label{performancedifference}
\end{figure}
\paragraph{Coordination Ability in Stationary and Non-stationary Settings} 
At first glance, we compare Fastap against the mentioned baselines to investigate the coordination ability under stationary and non-stationary conditions, as shown in Fig.~\ref{main_exp}. We can find all algorithms suffer from coordination ability degradation when teammates are in a non-stationary manner, indicating a specific consideration of teammates' policy sudden change in a non-stationary environment is needed.
When only using local information to obtain a context to capture the teammates' information, methods like PERAL and LIAM show indistinctive coordination improvement in stationary and non-stationary settings, PEARL performs even worse than vanilla QMIX, demonstrating that successful meta-learning approaches in SARL cannot be implemented without modification in the MARL setting. Furthermore, when learning a teammate's behavior context extraction model in both global and local ways, ODITS shows superior performance in the two mentioned conditions, manifesting the necessity of utilizing global states to improve training efficiency. Besides, ESCP also reveals a relatively better coordination capability, demonstrating the effectiveness of optimizing a context encoder with fast adaptability. 
Fastap achieves the best performance on all benchmarks both in stationary and non-stationary conditions, and suffers from the least performance degradation when tested in a non-stationary condition in most environments (see Fig.~\ref{performancedifference}), showing the effectiveness and high efficiency of the proposed method.  
\begin{table*}
\centering
\resizebox{\textwidth}{!}{
\begin{tabular}{l|ccccccc} 
\hline
                          $\mathcal{U}$   & Fastap       & Fastap\_wo\_CRP          & ODITS        & LIAM         & QMIX         & PEARL        &   ESCP\\ 
\hline
stationary                    & $\mathbf{0.642\pm0.008}$ & $0.594\pm0.015$ & $0.637\pm0.008$ & $0.597\pm0.029$ & $0.569\pm0.033$ & $0.507\pm0.021$ &   $0.618\pm0.040$   \\
{\cellcolor[rgb]{0.893,0.893,0.893}}$U[5, 8]$ & {\cellcolor[rgb]{0.893,0.893,0.893}}$\mathbf{0.562\pm0.012}$ & {\cellcolor[rgb]{0.893,0.893,0.893}}$0.400\pm0.020$ & {\cellcolor[rgb]{0.893,0.893,0.893}}$0.352\pm0.002$ & {\cellcolor[rgb]{0.893,0.893,0.893}}$0.415\pm0.026$ & {\cellcolor[rgb]{0.893,0.893,0.893}}$0.306\pm0.038$ & {\cellcolor[rgb]{0.893,0.893,0.893}} $0.288\pm0.019$& {\cellcolor[rgb]{0.893,0.893,0.893}} $0.404\pm0.026$    \\
$U[6, 7]$           & $\mathbf{0.567\pm0.001}$ & $0.444\pm0.314$ & $0.487\pm0.022$ & $0.454\pm0.157$ & $0.444\pm0.221$ & $0.333\pm0.000$  & $0.556\pm0.125$     \\
$U[2, 9]$           & $0.484\pm0.285$ & $0.222\pm0.133$ & $0.416\pm0.182$ & $0.401\pm0.078$ & $0.443\pm0.205$ & $0.205\pm0.114$ &   $\mathbf{0.514\pm0.314}$   \\
$U[3, 6]$           & $\mathbf{0.518\pm0.136}$ & $0.366\pm0.217$ & $0.444\pm0.314$ & $0.388\pm0.283$ & $0.353\pm0.272$ &  $0.264\pm0.066$ &   $0.502\pm0.120$   \\
$U[3, 3]$           & $\mathbf{0.384\pm0.272}$ & $0.246\pm0.141$ & $0.342\pm0.118$ & $0.362\pm0.208$ & $0.222\pm0.314$ & $0.243\pm0.172$ &   $0.271\pm0.157$   \\
\hline
\end{tabular}}
\caption{The final average return $\pm$ std in LBF, where $\mathcal{U}$ is the sudden change probability distribution of open Dec-POMDP that controls the frequency of sudden change, and $U[m, n]$ denotes a discrete uniform distribution parameterized by $m$ and $n$. The row of the original training sudden change distribution $\mathcal{U}=U[5, 8]$ is highlighted as \colorbox{lightgray}{gray}.}
\label{generalization_ood}
\end{table*}

\begin{figure}
\setlength{\abovecaptionskip}{0cm}
  \centering
  \subfigure[LBF]{
  \label{ablation_lbf}
      \includegraphics[width=0.226\textwidth]
      {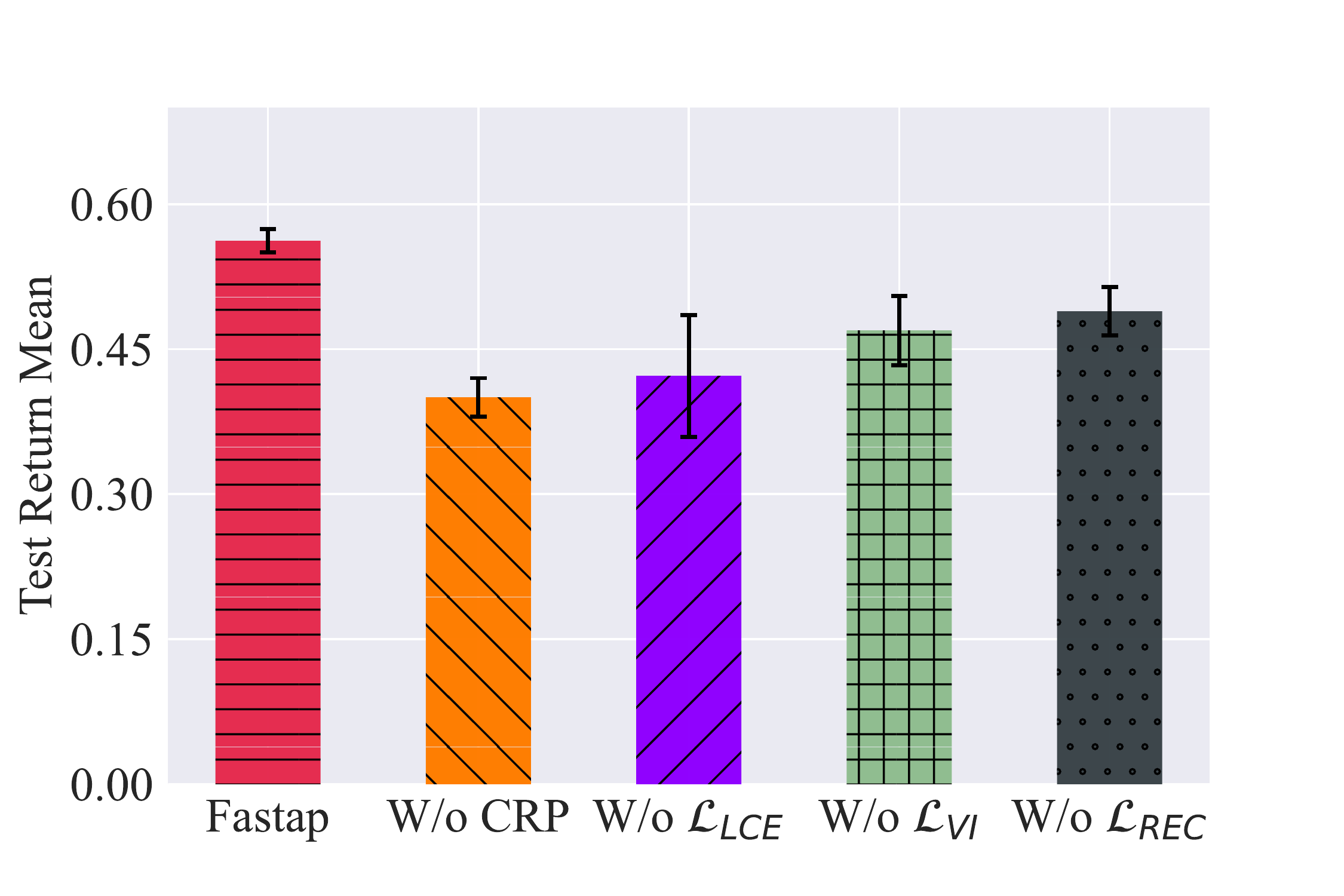}
  }
  \subfigure[PP]{
  \label{ablation_simpletag}
      \includegraphics[width=0.226\textwidth]
{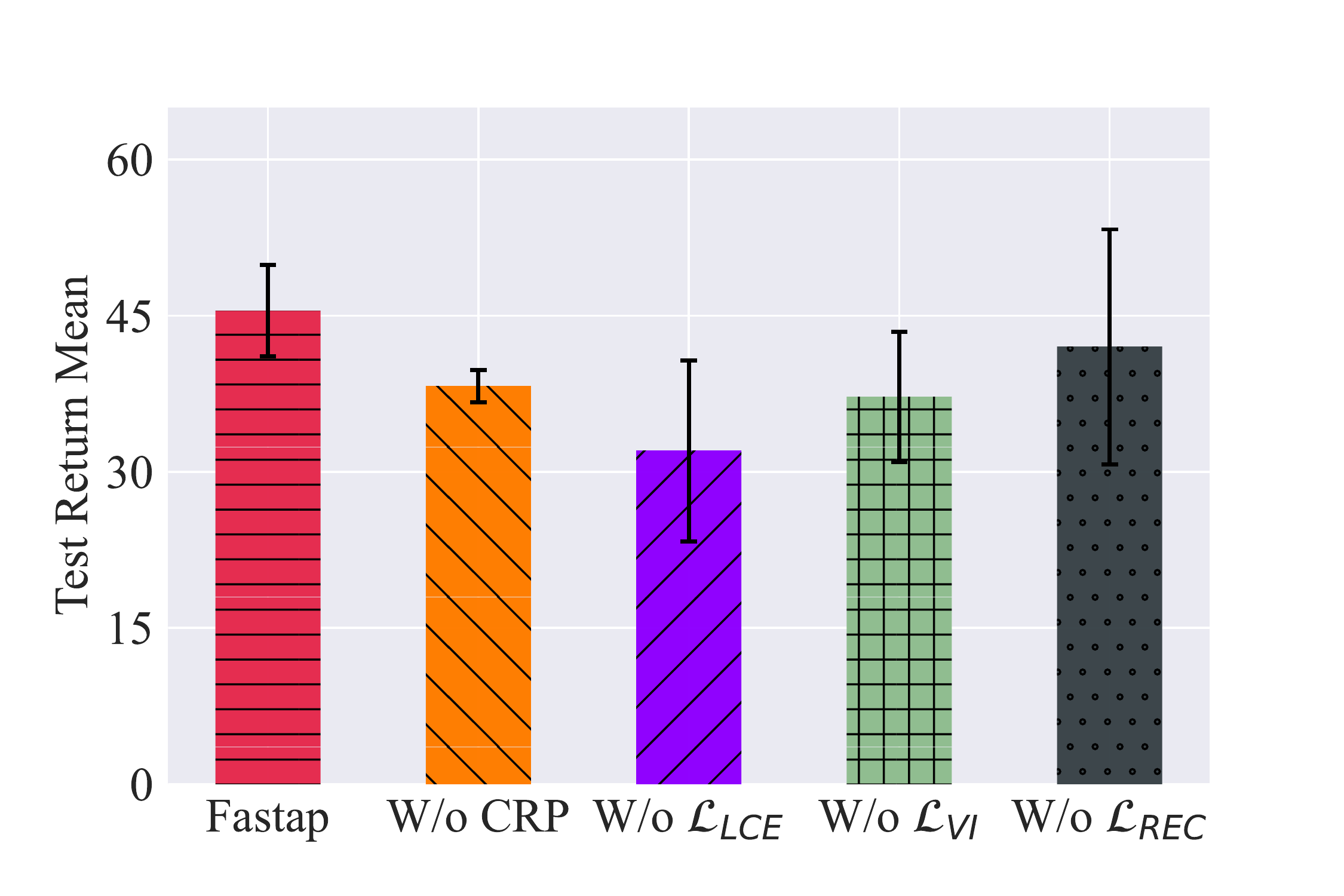}
  }
  \caption{Ablation Studies.}
  \label{ablation}
\end{figure}

\paragraph{Ablation Studies}
As Fastap is composed of multiple components,  we here design ablation studies on benchmarks LBF and PP to investigate how they impact the coordination performance of Fastap under non-stationary settings. First, for the infinite mixture model of dynamic teammate generation, we derive $\textit{W/o CRP}$ by removing the CRP process and taking each newly generated teammate group as a new cluster. Next, to explore whether a teammate-behavior-sensitive encoder helps improve adaptability, we introduce  $\textit{W/o LCE}$ by removing $\mathcal{L}_{\text{LCE}}$ of local encoders. Furthermore, we pick up $\textit{W/o MI}$ to investigate how maximizing mutual information between global and local contexts accelerates learning efficiency. Finally, $\textit{W/o REC}$ is introduced to check the impact of the auxiliary optimization objective that involves agent modeling. As is shown in Fig.~\ref{ablation}, $\textit{W/o CRP}$ and $\textit{W/o MI}$ suffer the most severe performance degradation in LBF and PP, respectively, manifesting the benefit of the introduction of CRP model and that teammate-behavior-sensitive encoders do help agents adapt to sudden change of teammates rapidly. Besides, when removing $\mathcal{L}_{\text{MI}}$, the performance gap $\textit{W/o MI}$  shows in two benchmarks demonstrate the necessity of utilizing global information to facilitate the learning of local context encoders. Finally, we also find agent modeling helps learn more informative context and brings about a slight coordination improvement.
\begin{figure}
  \centering 
  \label{hotmap}
    \includegraphics[height=28mm]{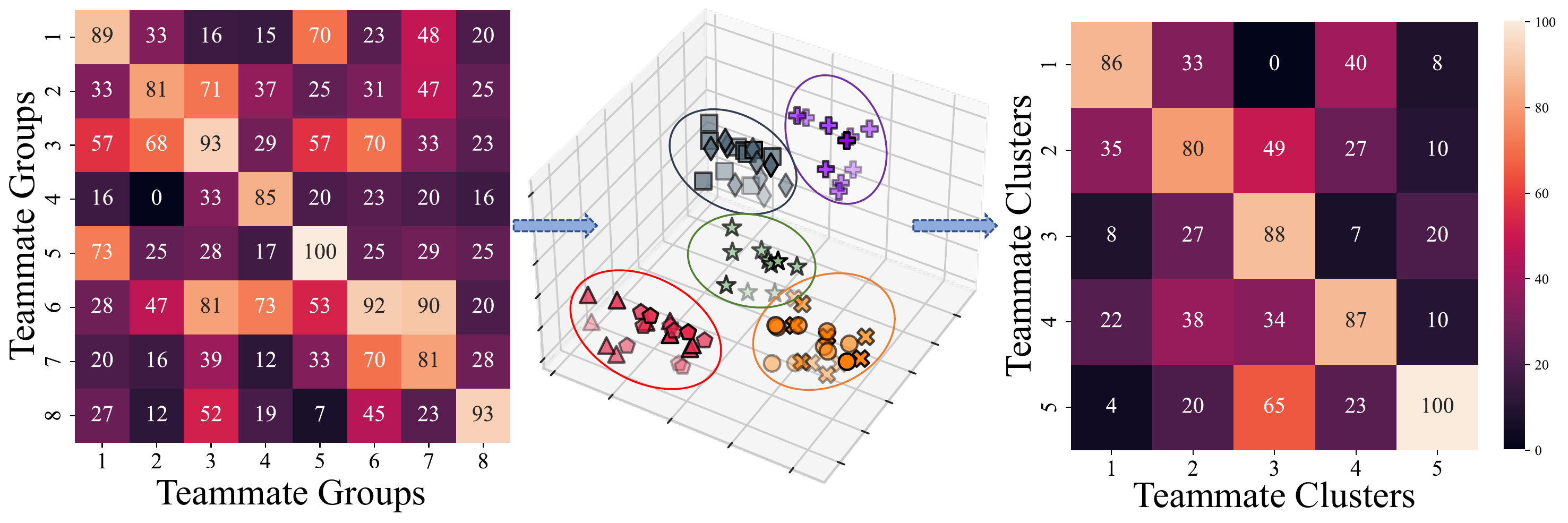}
    \caption{Cross-Play performance before and after CRP and teammate behavior embeddings.}
    \label{visual}
\end{figure}
\paragraph{Comparisons in (OOD) Non-stationary Setting.}
As this study considers a setting where the frequency of uncontrolled teammates' sudden change follows a fixed probability distribution $\mathcal{U}$, which is set to be a uniform distribution, we evaluate the generalization ability when altering the changing frequency during testing. The experiments on LBF are conducted with the distribution $\mathcal{U}=U[5, 8]$ during training. As shown in Tab.~\ref{generalization_ood}, we compare the final returns of different learned policies in LBF by altering the distribution $\mathcal{U}$. Although different approaches obtain similar coordination ability in stationary conditions, they suffer from strong performance degradation when altering teammates' policy-changing frequency (e.g., ODITS suffer from close to half performance degradation in sudden change[3, 3]).  On the other hand, Fastap and ESCP achieve outstanding generalization ability in both in-distribution and OOD settings mostly. 
More specifically, in the stationary setting, Fastap outperforms the best baseline ODITS by $0.005$, while in the original non-stationary setting, the gap increases to $0.147$. We also find Fastap shows inferiority to ESCP in setting sudden change[2, 9],  we believe that both methods fail to perform well under the 2-timestep sudden change interval, while Fastap sacrifices a part of the performance under large timestep sudden change interval that might happen in $U[2, 9]$. A more robust policy in diverse conditions would be developed in the future.

\subsection{Teammate Adaptation Analysis} \label{analysis}
Here we conduct experiments to investigate the CRP model and teammate adaptation progress. We first verify whether CRP helps acquire distinguishable boundaries of teammates' behaviors by performing Cross-Play~\citep{DBLP:conf/icml/HuLPF20} experiments on LBF before and after CRP. As shown in the left part of Fig.~\ref{visual}, for generation process of 8 teammate groups, we find that the values on the diagonal from the top left to the bottom right are relatively larger. However, several high performances of other points (e.g., Teammate groups 2 and 3) indicate that the generated teammate groups might share similar behavior. To help relieve the negative influence caused by taking teammate groups with similar behavior as two different types, CRP is applied to learn the behavior type and assign teammates with similar behavior to the same cluster. Further, we sample latent variables generated by $E_{\omega_1}(\tau_k)$ and reduce the dimensionality by principal component analysis (PCA)~\citep{wold1987principal}. We find that latent variables assigned to the same cluster (the ellipse) are distributed in the adjacent areas. Cross-Play experiments are also conducted on the teammate clusters after CRP, and we find from the right part of Fig.~\ref{visual} that teammates belonging to different clusters achieve low performance when paired together, indicating the effectiveness of CRP.

\begin{figure}
  \centering
     \label{visual_context}
    \subfigure[Snapshot]{
        \label{snapshot}
        \includegraphics[height=34.5mm]{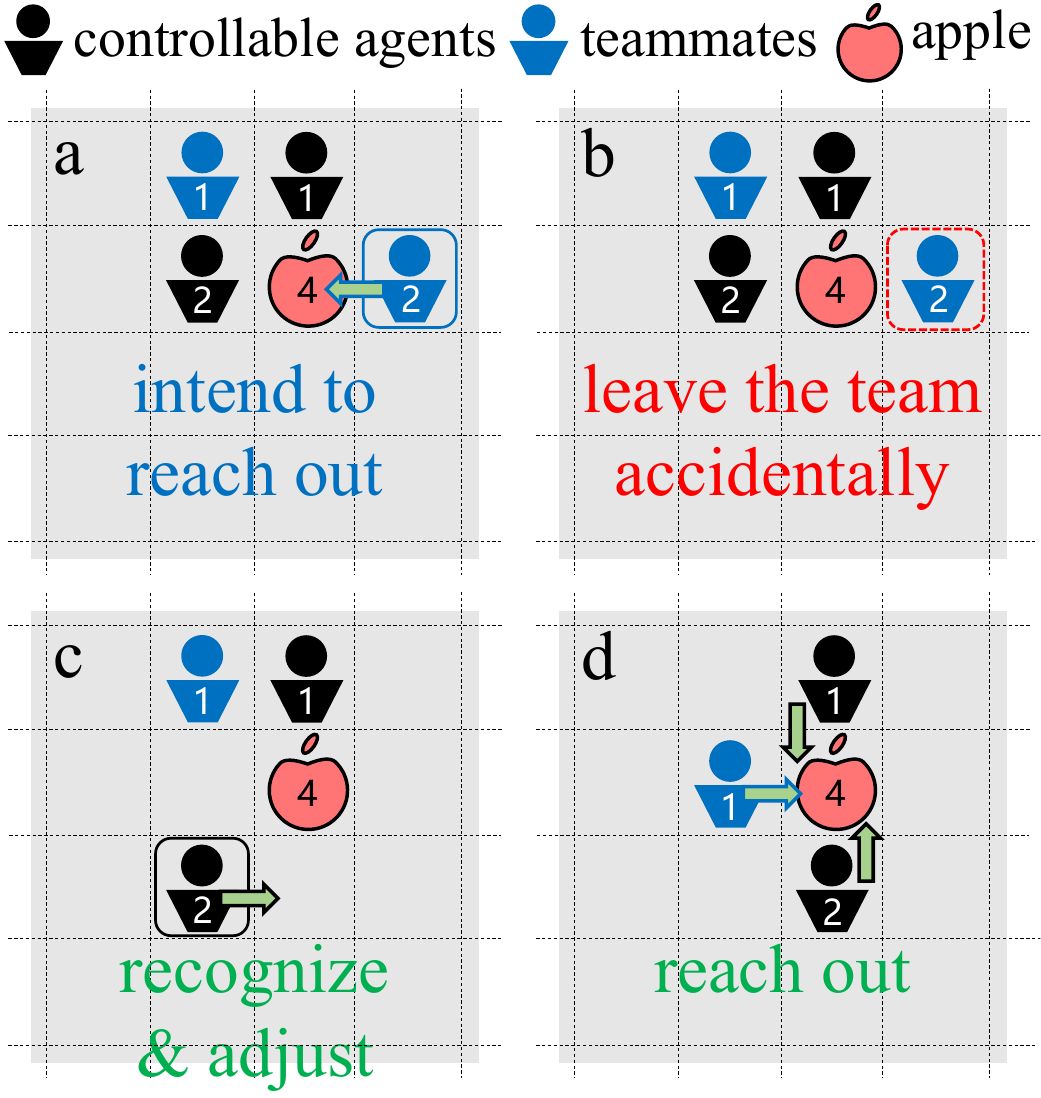}
        
    }
    \subfigure[Context Curve]
    {\label{context_curve}
    \includegraphics[height=32.5mm]{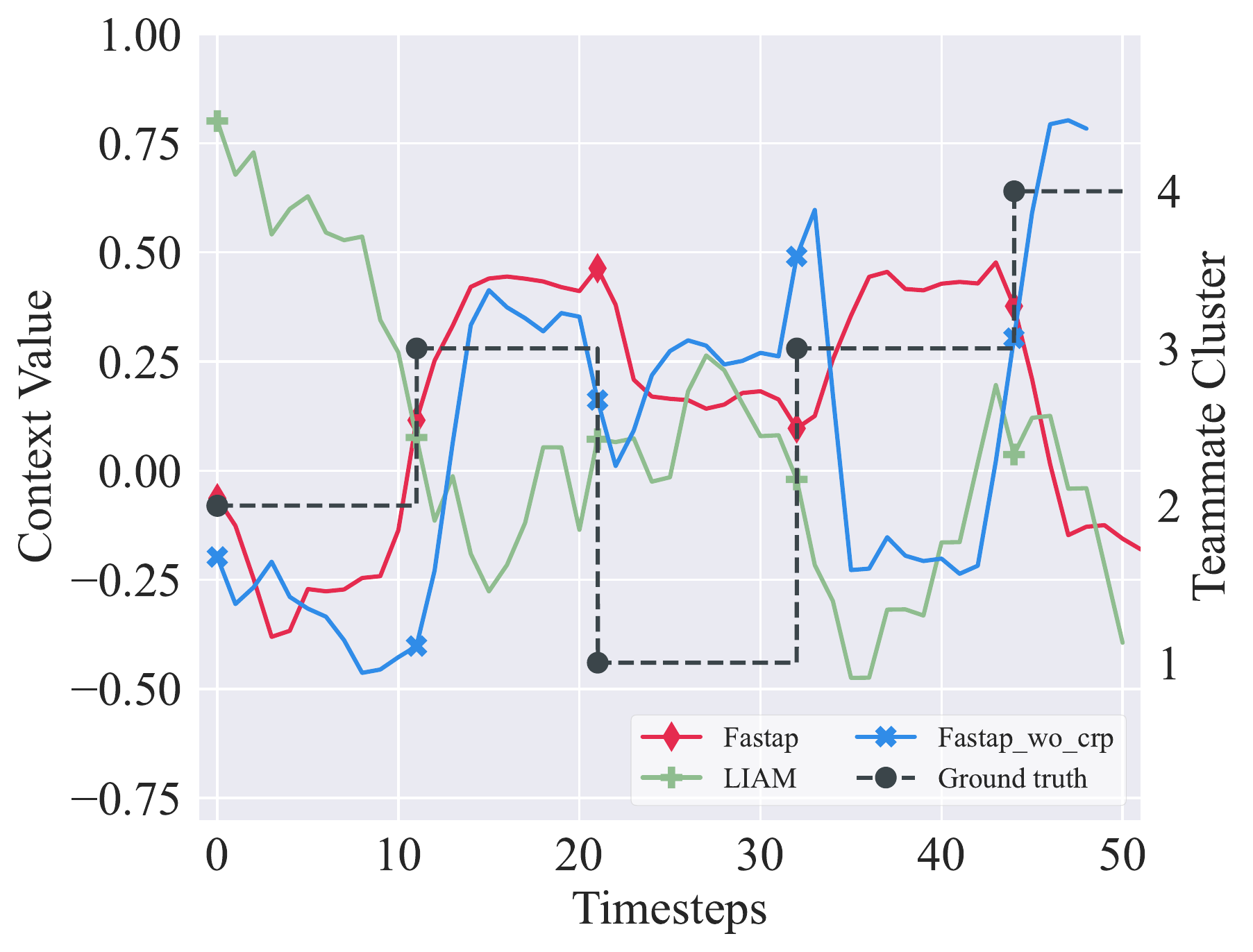}}

    \caption{Teammate adaptation visualization.}
\end{figure}

To investigate how teammate-behavior-sensitive encoders help adapt to teammates' sudden change rapidly, we also visualize the fragment snapshot of an episode during testing as shown in Fig.~\ref{snapshot}. When a teammate and two controlled agents are trying to reach out for an apple and win the score as they were intended, the teammate accidentally leaves out the team, and they fail to get the reward provisionally. However, the controlled agents learned by Fastap recognize the situation and switch out the policy rapidly by moving downward and coordinating with the other teammate to attain the reward. Meanwhile, we record the latent context vector in different timesteps of one episode. Fastap encodes the context to four-dimensional vectors in LBF, and we reduce the dimensionality to one-dimensional scalars by PCA. We scatter the points in Fig.~\ref{context_curve} together with the contexts learned by LIAM and ablation Fastap\_wo\_CRP. The results imply that the contexts learned by Fastap are sensitive to the sudden change of teammates, and when the teammates are stable, the latent context is stable and flat. Despite the fact that agent modeling helps recognize the teammates' behavior, the context curve of LIAM is still hysteretic and unstable. Meanwhile, the ablation Fastap\_wo\_CRP can also adapt to new teammates rapidly, but it fails to recognize the teammates with similar behavior and results in the unstable latent context (e.g., Teammate Cluster 3).


\subsection{Transfer and Sensitive Studies} \label{bonus}
\begin{figure}
\setlength{\abovecaptionskip}{0cm}
  \centering
  \subfigure[LBF]{
  \label{transfer_lbf}
      \includegraphics[width=0.222\textwidth]
      {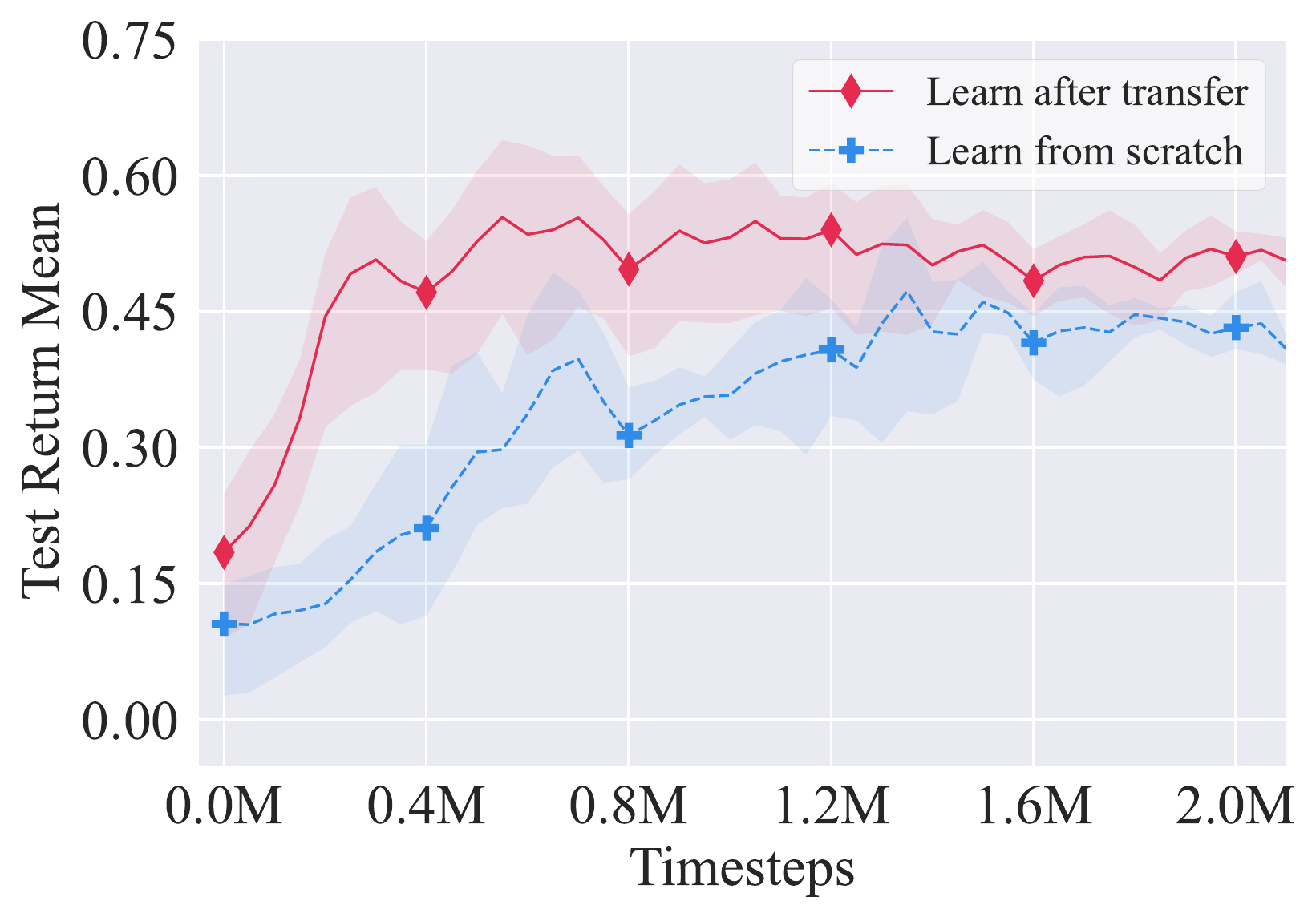}
  }
  \subfigure[PP]{
  \label{transfer_simpletag}
      \includegraphics[width=0.222\textwidth]
{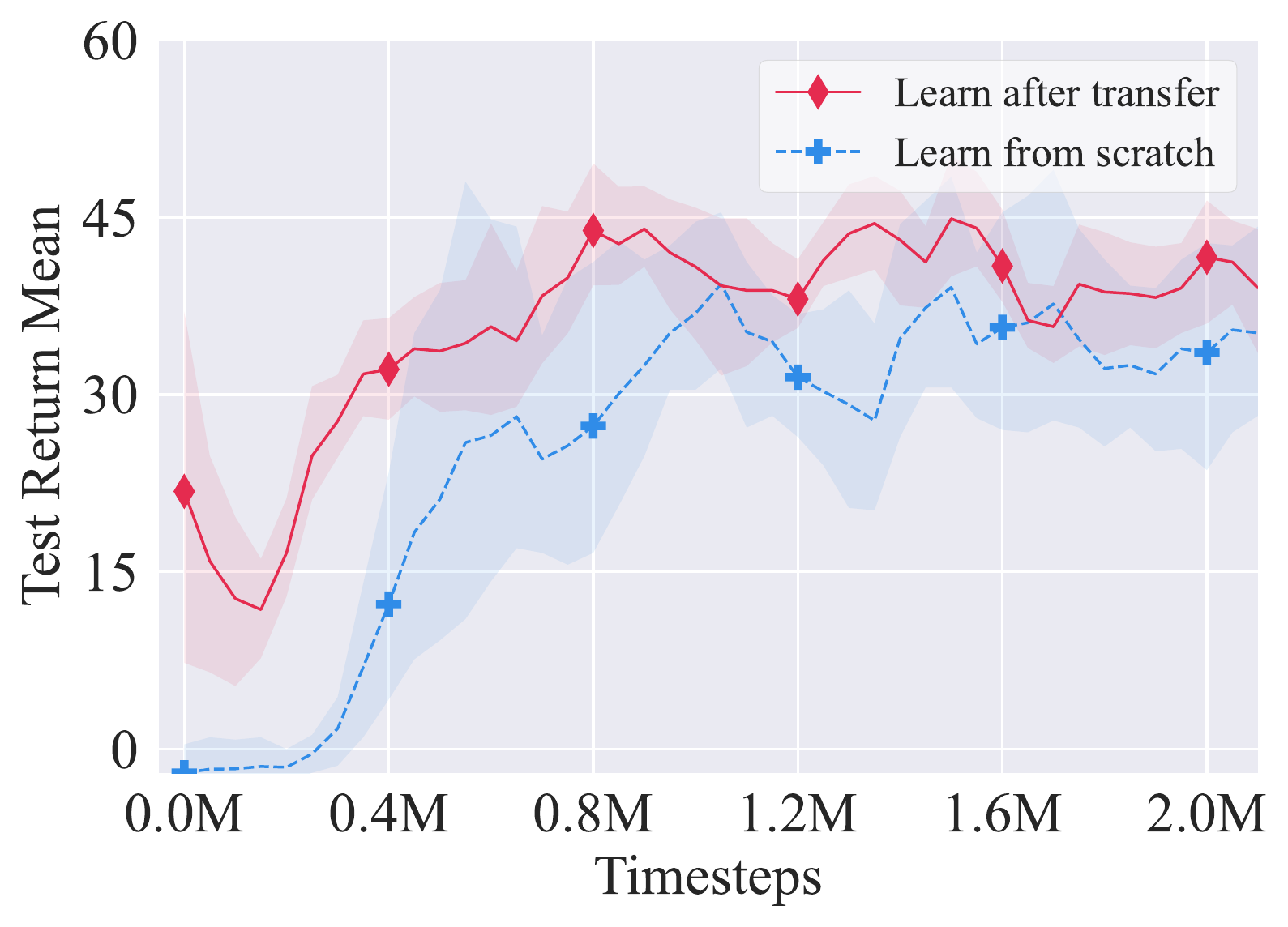}
  }
  \caption{Policy Transfer Ability.}
  \label{transfer}
\end{figure}

Our Fastap learns teammates recognition module to cope with teammates that might change suddenly in one episode. The sudden change distribution $\mathcal{U}$ that controls the frequency of changing is fixed, and a more frequent change or a larger gap of waiting interval tends to make the training more difficult. Here, we investigate the policy transfer ability of Fastap by comparing the performance after fine-tuning and learning from scratch. Concretely, we train Fastap agents under the sudden change distribution $\mathcal{U}_{\text{source}}=U[5, 8]$ for $0.6$M timesteps and initialize the trained network with the saved checkpoint under the target setting with $\mathcal{U}=\mathcal{U}_{\text{target}}=U[3, 6]$. The learning curves demonstrated in Fig.~\ref{transfer} show that agents trained under $\mathcal{U}_{\text{source}}$ possess a jumpstart compared with the random initialization, and we hope it could accelerate the learning in a new environment by reusing previously learned knowledge.

\begin{figure}
\setlength{\abovecaptionskip}{0cm}
  \centering
  \subfigure[Sensitivity of $\alpha_{\text{GCE}}$]{
  \label{sensitivity_gce}
      \includegraphics[width=0.226\textwidth]
      {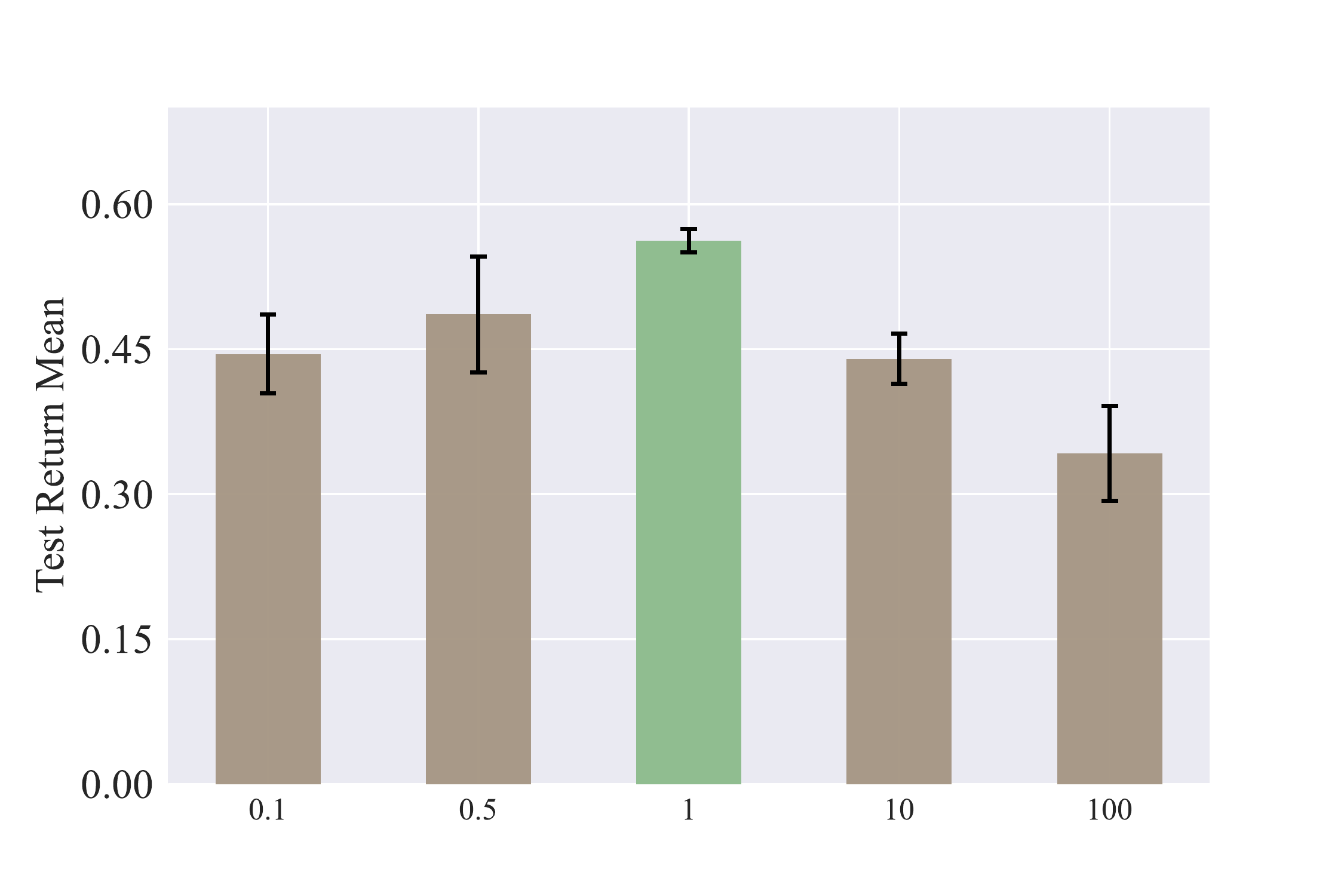}
  }
  \subfigure[Sensitivity of $\alpha_{\text{MI}}$]{
  \label{sensitivity_mi}
      \includegraphics[width=0.226\textwidth]
{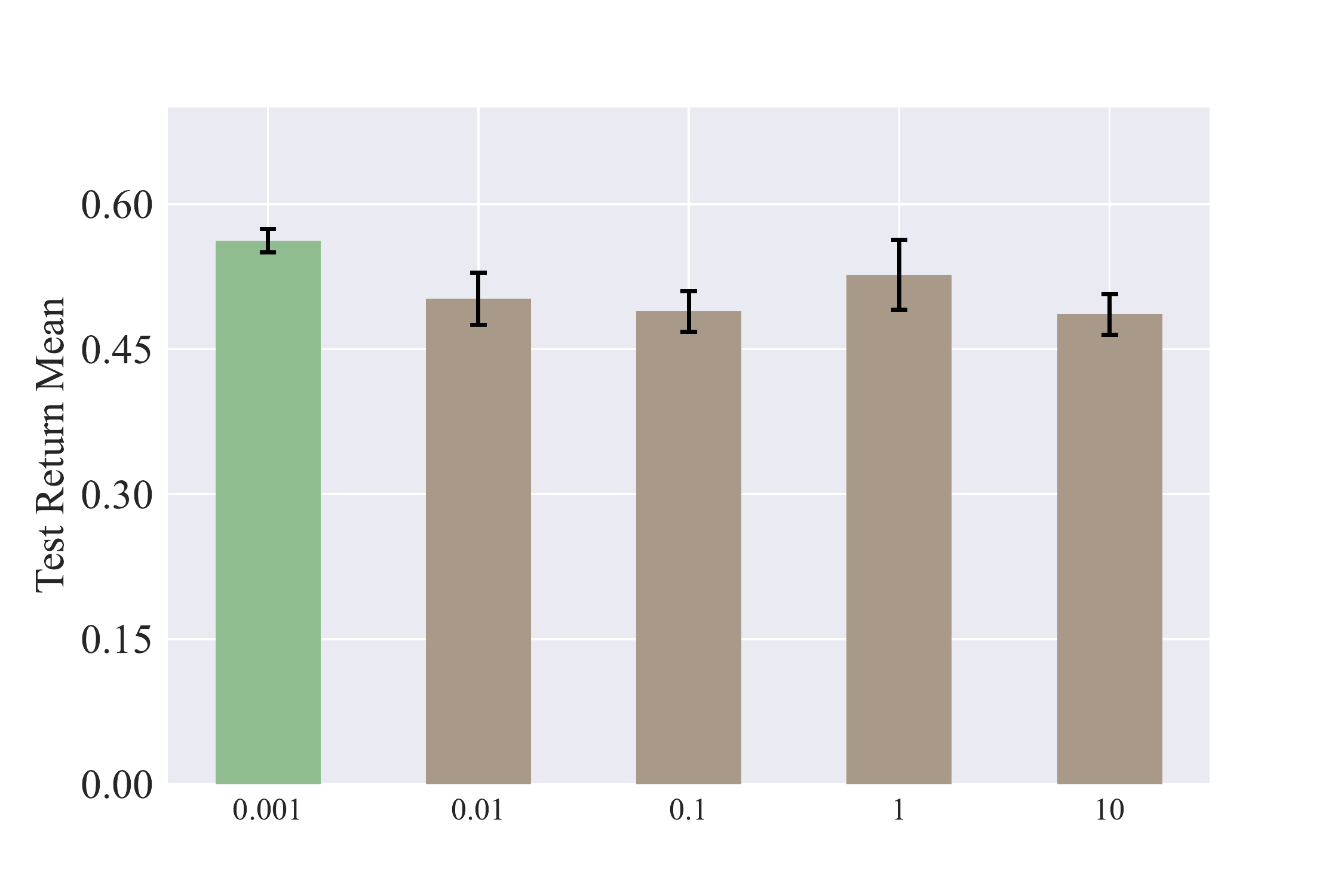}
  }
  \subfigure[Sensitivity of $\alpha_{\text{LCE}}$]{
  \label{sensitivity_lce}
      \includegraphics[width=0.226\textwidth]
{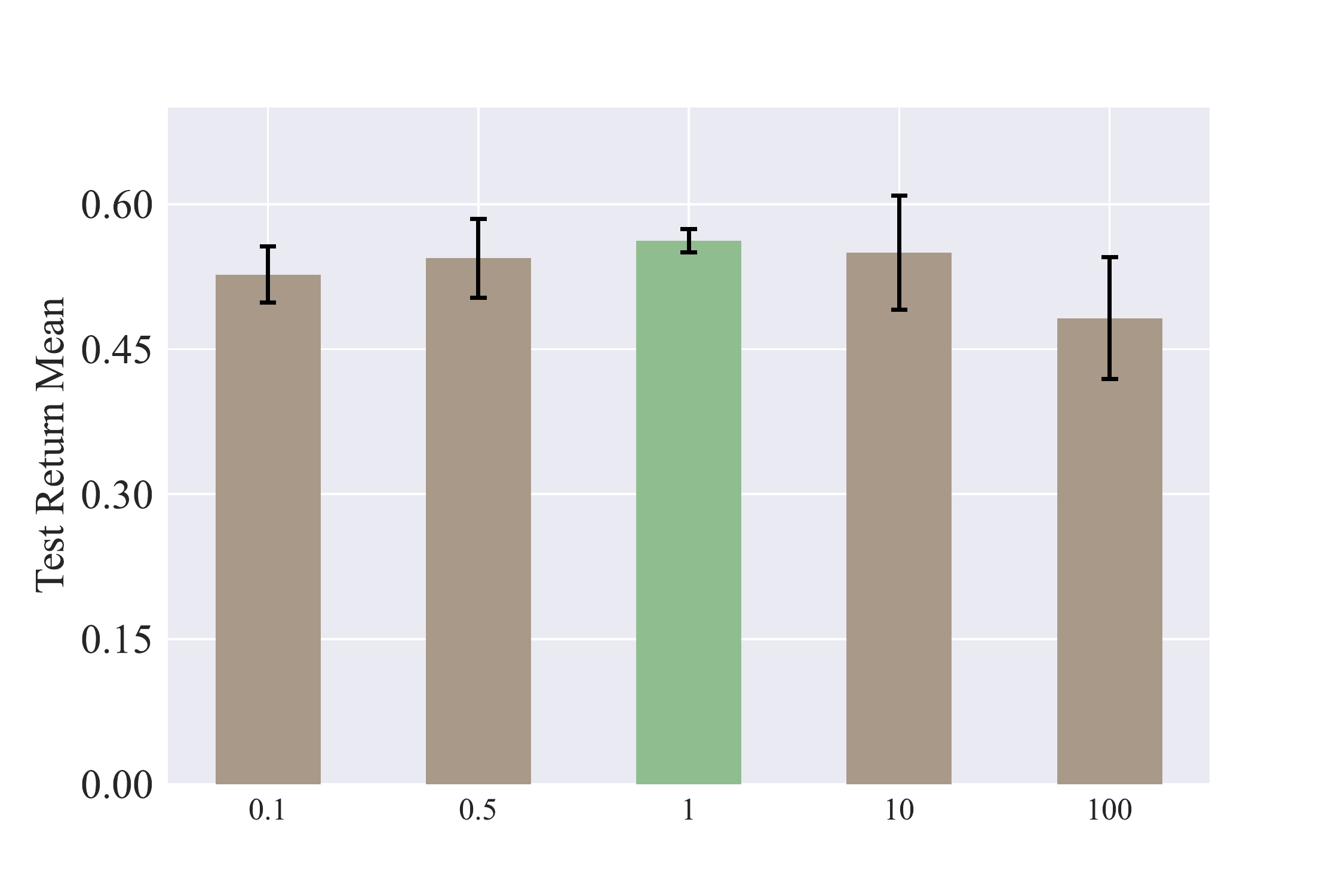}
  }
  \subfigure[Sensitivity of $ \alpha_{\text{REC}}$]{
  \label{sensitivity_rec}
      \includegraphics[width=0.226\textwidth]
{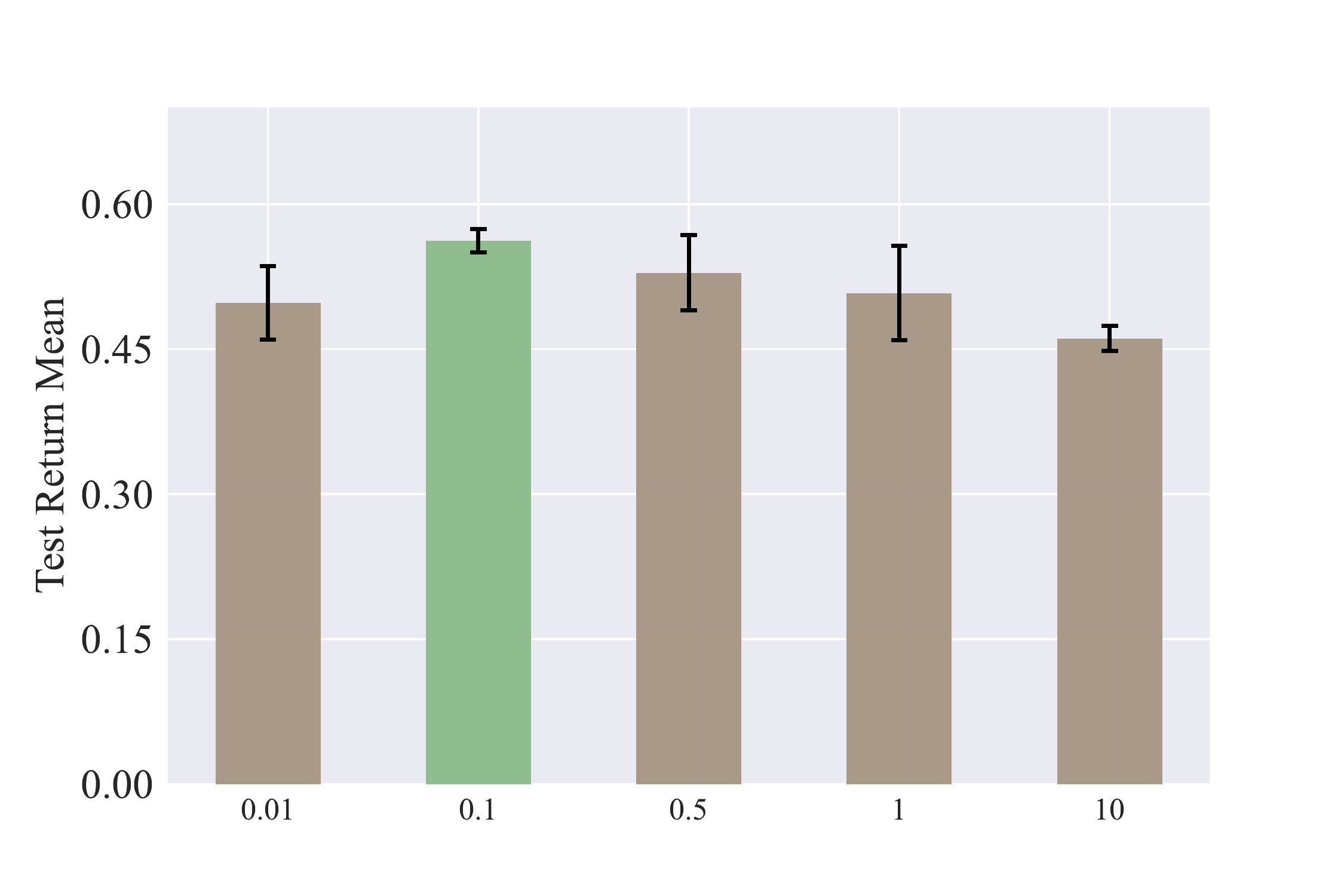}
  }
  \caption{Sensitivity Studies on LBF.}
  \label{sensitivity}
\end{figure}

As Fastap includes multiple hyperparameters, here we conduct experiments on benchmark  LBF to investigate how each one influences the coordination ability. First, $\alpha_{\text{GCE}}$ balances the trade-off between the TD-loss and the global context optimization object. If it is too small, agents may coordinate in stationary environment excessively, ignoring the extraction of teammates context information. On the other hand, if it is too large, agents pay much attention to teammates identification with risk of overfitting to specific teammates types. We thus find each hyperparameter via grid-search.  As shown in Fig.~\ref{sensitivity_gce}, we can find that $\alpha_{\text{GCE}}=1$ is the best choice in this benchmark. 
$\alpha_{\text{MI}}$ influences the optimization of local encoder $f_{\phi_i}$ to recognize the team situation.  Fig.~\ref{sensitivity_mi} shows that $\alpha_{\text{cont}_g}=0.001$ performs the best. wW can find that $\alpha_{\text{LCE}}=1, \alpha_{\text{REC}}=0.1$ are the corresponding best choices in a similar way.

\section{Final Remarks}
In this work,  we study the teammates' adaptation problem when some coordinators suffer from the sudden policy change.
 We first formalize this problem as an open Dec-POMDP, where some coordinators from a team may sustain policy changes unpredictably within one episode, and we train multiple controlled agents to adapt to this change rapidly. For this goal, we propose Fastap, an efficient approach to learn a robust multi-agent coordination policy by capturing the teammates' policy-changing information. Extensive experimental results on stationary and non-stationary conditions from different benchmarks verify the effectiveness of Fastap, and more analysis results also confirm it from multiple aspects. Our method can be seen as a primary attempt for the open-environment setting~\citep{zhou2022open} in cooperative MARL, and we sincerely hope it can be a solid foothold for applying MARL to practical applications. For future work, researches on the changing of action/observation space of the MARL system or utilizing techniques like transformer~\citep{DBLP:conf/nips/VaswaniSPUJGKP17} to obtain a generalist coordination policy for non-stationary from diverse sources and degrees is of great value.  

{\Large \textbf{Appendix}}

\section{Related Work} 
\paragraph{Cooperative Multi-agent Reinforcement Learning}
Many real-world problems are made up of multiple interactive agents, which could usually be modeled as a multi-agent system~\citep{DBLP:journals/access/DorriKJ18}. Among the multitudinous solutions, Multi-Agent Reinforcement Learning (MARL)~\citep{zhang2021multi} has made great success profit from the powerful problem-solving ability of deep reinforcement learning~\citep{ Wang2020DeepRL}. Further, when the agents hold a shared goal, this problem refers to cooperative MARL~\citep{oroojlooy2022review}, showing great progress in diverse domains like path finding~\citep{sartoretti2019primal},  active voltage control~\citep{DBLP:conf/nips/WangXGSG21}, and dynamic algorithm configuration~\citep{xue2022multiagent}, etc. Many methods are proposed to facilitate coordination among agents, including policy-based ones (e.g., MADDPG~\citep{maddpg}, MAPPO~\citep{mappo}),  value-based series like VDN~\citep{vdn}, QMIX~\citep{qmix}, or other techniques like transformer~\citep{wen2022multiagent} and many variants~\citep{gorsane2022towards}, demonstrating remarkable coordination ability in a wide range of tasks like
SMAC~\citep{pymarl}, Hanabi~\citep{mappo}, GRF~\citep{wen2022multiagent}. 
Besides the mentioned approaches and the corresponding variants, many other methods are also proposed to investigate the cooperative MARL from other aspects, including casual inference among agents~\citep{grimbly2021causal}, policy deployment in an offline way for real-world application~\citep{DBLP:conf/nips/YangMLZZHYZ21}, 
 communication~\citep{zhu2022survey} for partial observability, model learning for sample efficiency improvement~\citep{wang2022model}, policy robustness when perturbations occur~\citep{guo2022towards,ma3c}, training paradigm like CTDE (centralized training with decentralized execution)~\citep{DBLP:conf/atal/LyuXDA21}, testbed design for continual coordination validation~\citep{DBLP:conf/icml/NekoeiBCC21}, and ad hoc teamwork~\citep{mirsky2022survey}, offline learning in MARL~\cite{guan2023efficient,zhang2023discovering}, etc.

\textbf{Non-stationary} is a longstanding topic in single-agent reinforcement learning (SARL)~\citep{Padakandla2019ReinforcementLA,Padakandla2020ASO}, where the environment dynamic (e.g., transition and reward functions) of a learning system may change over time. For SARL, most existing works focus on inter-episode non-stationarity, where decision
processes are non-stationary across episodes, including multi-task setting~\citep{VithayathilVarghese2020ASO}, continual reinforcement learning~\citep{DBLP:journals/jair/KhetarpalRRP22}, meta reinforcement learning~\citep{beck2023survey}, etc., these problems can be formulated as a contextual MDP~\citep{hallak2015contextual}, and could be solved by techniques like task embeddings learning. Other works also consider intra-episode non-stationarity, where an agent may suffer from dynamic drifting within one single episode~\citep{DBLP:conf/rss/KumarFPM21,DBLP:conf/iclr/RenSJSWB22,chen2022an,escp,DBLP:conf/case/DastiderL22,feng2022factored}. Specifically, HDP-C-MDP~\citep{DBLP:conf/iclr/RenSJSWB22} assumes the latent context to be finite and Markovian, and adapts a sticky Hierarchical Dirichlet Process (HDP) prior for
model learning; while FANS-RL~\citep{feng2022factored} assumes the latent context is Markovian and the environment can be modeled as a factored MDP; ESCP~\citep{escp} considers the sudden changes one agent may encounter and obtains a robust policy via learning an auxiliary context recognition model. Experiments show that in environments with both in-distribution and out-of-distribution
parameter changes, ESCP can not only better recover the environment encoding, but also adapt more rapidly to the post-change environment
; SeCBAD~\citep{chen2022an} further assumes the environment context usually stays stable for a stochastic
period and then changes in an abrupt and unpredictable manner. Linda~\cite{cao2021linda} learns to decompose local information and build awareness for
each teammate, which promotes coordination ability in multiple environments.

\textbf{Open Multi-agent System} considers the problem where agents may join or leave while the process is ongoing, causing the system's composition and size to evolve over time~\cite{hendrickx2017open}. In previous works, the multi-agent problem has mainly been modeled for planning, resulting in various problem formulations such as Open Dec-POMDP~\cite{Cohen2017OpenDP}, Team-POMDP~\cite{Cohen2018MonteCarloPF,Cohen2019PowerIF}, I-POMDP-Lite~\cite{Chandrasekaran2016IndividualPI,Eck2019ScalableDP}, CI-POMDP~\cite{Kakarlapudi2022DecisiontheoreticPW}, and others.  Recently, some works consider the open multi-agent reinforcement learning problems. GPL~\cite{rahman2021towards} formulates the Open Ad-hoc Teamwork as OSBG and assumes global observability for efficiency, which may be hard to achieve in the real world. Additionally, it uses a GNN-based method that works only on the single controllable agent setting and is not scalable enough to be extended to multiple controllable agents setting. ROMANCE~\cite{romance} models the problem where the policy perturbation issue when testing in a different environment as a limited policy adversary Dec-POMDP (LPA-Dec-POMDP), and then proposes \textbf{Ro}bust \textbf{M}ulti-\textbf{A}ge\textbf{n}t \textbf{C}oordination via \textbf{E}volutionary Generation of Auxiliary Adversarial Attackers (ROMANCE), which enables the trained policy to encounter diversified and strong auxiliary adversarial attacks during training, thus achieving high robustness under various policy perturbations. 

 
Different from the SARL setting, non-stationarity is an inherent challenge for MARL, as the agent's policy may be instability caused by the concurrent learning of multiple policies of other agents~\citep{papoudakis2019dealing}. Previous works mainly focus on solving the non-stationary in the training phase, using techniques like agent modeling~\citep{DBLP:journals/ai/AlbrechtS18}, meta policy adaptation~\citep{DBLP:conf/icml/Kim0RSAHLTH21}, experience sharing~\citep{DBLP:conf/nips/ChristianosSA20}.
Other works concentrate on non-stationarity across episodes, Previous works have focused on solving non-stationarity in the training phase using techniques such as multi-task training~\citep{qin2022multi},  training policy for zero-shot coordination~\citep{DBLP:conf/icml/HuLPF20}. 
Despite the progress made by these approaches, they do not address non-stationarity caused by teammates' policy sudden changes, which is a crucial and urgent need. As for the open MARL, our work takes a different perspective by emphasizing the general coordination and fast adaptation ability of learned controllable agents in the context of MARL.


\section{Details about Derivation}

\subsection{Details about CRP and derivation of cluster assignment}
Chinese restaurant process (CRP)~\citep{crp} is a discrete-time stochastic process that defines a prior distribution over the cluster structures, which can be described simply as follows. A customer comes into a Chinese restaurant, he chooses to sit down alone at a new table with a probability proportional to a concentration parameter $\alpha$  or sits with other customers with a probability proportional to the number of customers sitting on the occupied table. Customers sitting at the same table will be assigned to the same cluster. Concretely, suppose that $K$ customers sit in the restaurant currently. Let $z_i$ be an indicator variable that tells which table that $i^{\text{th}}$ sits on, and $n_m$ denote the number of customers sitting at the $m^{\text{th}}$ table, and $M$ be the total number of non-empty tables. Note that $\sum_{m=1}^M n_m=K$. The probability that the $K+1^{\text{th}}$ customer sits at the $m^{\text{th}}$ table is:
\begin{equation}
    \begin{aligned}
        P(z_{K+1}=m|\alpha)=\frac{n_m}{K+\alpha}, \quad m=1,...,M.
    \end{aligned}
\end{equation}
 There is some probability that the customer decides to sit at a new table and if the label of the new table is $M+1$, then:
\begin{equation}
    \begin{aligned}
        P(z_{K+1}=M+1|\alpha)=\frac{\alpha}{K+\alpha}.
    \end{aligned}
\end{equation}
Taken together, the two equations characterize the CRP.

 The cluster assignment of the $k^{\text{th}}$ generated teammate group $P(v_k^{(m)}|\tau_k^S,\tau_k^A)$ can be decomposed:
 \begin{equation}
     \begin{aligned}
         &P(v_k^{(m)}|\tau_k^S,\tau_k^A)\\ 
         &\quad=\frac{P(v_k^{(m}, \tau_k^S, \tau_k^A)}{P(\tau_k^S, \tau_k^A)}\\
         &\quad=\frac{P(\tau_k^S,\tau_k^A|v_k^{(m)})P(v_k^{(m)})}{P(\tau_k^S,\tau_k^A)}\\
         &\quad=\frac{P(\tau_k^A|\tau_k^S, v_k^{(m)})P(\tau_k^S|v_k^{(m)})P(v_k^{(m)})}{P(\tau_k^S,\tau_k^A)}\\
         &\quad\propto P(\tau_k^A|\tau_k^S, v_k^{(m)})P(\tau_k^S|v_k^{(m)})P(v_k^{(m)}).
     \end{aligned}
 \end{equation}
 As $\tau_k^S$ is a set of states that is not determined by the behavioral type of the teammates if neglecting the correlation in time dimensionality. $P(\tau_k^S|v_k^{(m)})$ can be considered as a constant. Accordingly, we would derive that $P(v_k^{(m)}|\tau^S_k, \tau^A_k)\propto P(v_k^{(m)})P(\tau_k^A|\tau_k^S; v_k^{(m)})$.
 
\subsection{The full derivation of $\mathcal{L}_{\text{GCE}}$}
To guide the context encoder to identify and track the sudden change rapidly, ESCP~\citep{escp} proposes the following optimization objective:
\begin{equation}
    \begin{aligned}
        \mathcal{L}_{\text{GCE}} = \sum_{m=1}^M\mathbb{E}[||z^m_t-\mathbb{E}[z^m_t]||_2^2] + ||\mathbb{E}[z^m_t] - u^m||_2^2,
    \end{aligned}
    \label{obj_escp}
\end{equation}
where $z_t^m$ is the representation that context encoder embeds in the $m^{\text{th}}$ environment,  $u^m$ is the oracle latent context vector, and $M$ is the number of environments. For a better understanding, we would explain the meanings of symbols based on our setting in the following. So, $z_t^m$ is the latent context vector when paired with teammates belonging to the $m^{\text{th}}$ cluster, and $u^m$ is the oracle behavior type.

 Since we have no access to the oracle $u^m$, a set of surrogates that possesses large diversity is required to be separable and representative. Meanwhile, $u^m$ is an intermediate variable used to guide $\mathbb{E}[z_t^m]$, so we could directly maximize the diversity of $\{\mathbb{E}[z_t^m]\}_{m=1}^M$ by maximizing the determinant of a relational matrix $R_{\{\mathbb{E}[z_t^m]\}}$. Each element of the relational matrix is:
\begin{equation}
    \begin{aligned}
            R_{\{\mathbb{E}[z_t^m]\}}(i, j) = \exp(-\kappa{||\mathbb{E}[z_t^i]-\mathbb{E}[z_t^j]||_2^2}),
    \end{aligned}
\end{equation}
where $\kappa$ is the radius hyperparameter of the radius basis function applied to calculate the distance of two vectors. The objective function can now be written as:
\begin{equation}
    \begin{aligned}
        \mathcal{L}_{\text{GCE}} = \sum_{m=1}^M\mathbb{E}[||z^m_t-\mathbb{E}[z^m_t]||_2^2]-\log\det(R_{\{\mathbb{E}[z_t^m]\}}).
    \end{aligned}
\end{equation}

To stabilize the training process, ESCP substitutes $\mathbb{E}[z_t^m]$ with $\bar z^m$, which is the moving average of all past context vectors. $\{\bar z^m\}$ will be updated after sampling a new batch of $z_t^m$: 
\begin{equation}
    \begin{aligned}
        \bar z^m=\eta \text{sg}(\bar z^m)+(1-\eta)\mathbb E[z_t^m],
    \end{aligned}
\end{equation}
where $\text{sg}(\cdot)$ denotes stopping gradient, and $\eta$ is a hyperparameter controlling the moving average horizon.
\subsection {Variational Bound of teammates context approximation}
\label{mi}

In order to make context vector $e_t^{m, i}$ generated by local trajectory encoder $f_{\phi_i}$ informatively consistent with global context $z_t^m$ encoded by $g_{\theta}$, we propose to maximize the mutual information between $e_t^{m, i}$ and $z_t^m$ conditioned on the agent $i$'s local trajectory $\tau_t^{m, i}$. We draw the idea from variational inference~\citep{DBLP:conf/iclr/AlemiFD017} and derive a lower bound of this mutual information term.

\begin{theorem} 
Let $\mathcal{I}(e_{t}^{m, i};z_t^m
|\tau_t^{m,i})$ be the mutual information between the local context $e_t^{m, i}$ of agent $i$ and global context $z_t^m$ conditioned on agent $i$'s local trajectory $\tau^{m, i}_t$. The lower bound is given by
\begin{equation}
    \mathbb{E}_{\mathcal{D}}[\log q_{\xi}(e_{t}^{m, i}|z_t^m, \tau^{m, i}_t)]+\mathcal{H}(e_{t}^{m, i}|\tau^{m, i}_t).
\end{equation}
\end{theorem}
Here $m$ is the cluster id of the teammates cooperating with controlled agents to finish the task in this episode.

\begin{proof}
By a variational distribution $q_{\xi}(e_{t}^{m, i}|z_t^m, \tau^{m, i}_t)$ parameterized by $\xi$, we have
\begin{equation}
    \begin{aligned}
        &\mathcal{I}(e_{t}^{m, i};z_t^m
|\tau_t^{m,i})  \\
        =& \mathbb{E}_{\mathcal{D}}\Big[ \log \frac{p(e_{t}^{m, i};z_t^m
|\tau_t^{m,i})}{p(e_{t}^{m,i}|\tau_t^{m, i})p(z_t^m|\tau_t^{m, i})}\Big] \\
        =&\mathbb{E}_{\mathcal{D}}\left[\log \frac{p(e_{t}^{m, i}|z_t^m; \tau_t^{m,i})}{p(e_{t}^{m,i}|\tau_t^{m, i})}\right] \\
        =&\mathbb{E}_{\mathcal{D}}\left[\log \frac{q_{\xi}(e_{t}^{m, i}|z_t^m, \tau^{m, i}_t)}{p(e_{t}^{m,i}|\tau_t^{m, i})}\right]+ \\
        &D_{\text{KL}}(p(e_t^{m, i}|z_t^m, \tau_t^{m,i}) || q_{\xi}(e_{t}^{m, i}|z_t^m, \tau^{m, i}_t)) \\
        \geq & \mathbb{E}_{\mathcal{D}}\left[\log \frac{q_{\xi}(e_{t}^{m, i}|z_t^m, \tau^{m, i}_t)}{p(e_{t}^{m,i}|\tau_t^{m, i})}\right] \\
        = & \mathbb{E}_{\mathcal{D}}[\log q_{\xi}(e_{t}^{m, i}|z_t^m, \tau^{m, i}_t)]+\mathcal{H}(e_{t}^{m, i}|\tau^{m, i}_t).
    \end{aligned}
\end{equation}
\end{proof}
\section{Details About Baselines and Benchmarks}
\subsection{Baselines Used}
\paragraph{QMIX~\citep{qmix}:} As we investigate the integrative abilities of Fastap in the manuscript, here we introduce the value-based method QMIX~\citep{qmix} used in this paper. Our proposed framework Fastap follows the \textit{Centralized Training with Decentralized Execution} (CTDE) paradigm used in value-based MARL methods, as well as the Individual-Global-Max (IGM)~\citep{QTRAN} principle, which asserts the consistency between joint and local greedy action selections by the joint value function $Q_{\rm tot}(\boldsymbol{\tau}, \boldsymbol{a})$ and individual value functions $\left[Q_i(\tau^i, a^i)\right]_{i=1}^n$:
\begin{equation}
\begin{aligned}
  & \forall \boldsymbol{\tau} \in \boldsymbol{\mathcal{T}}, \underset{\boldsymbol{a} \in \boldsymbol{\mathcal{A}}}{\arg \max } Q_{\rm tot}(\boldsymbol{\tau}, \boldsymbol{a})= \\
     &\left(\underset{a^{1} \in \mathcal{A}}{\arg \max } Q_{1}\left(\tau^{1}, a^{1}\right), \ldots, \underset{a^{n} \in \mathcal{A}}{\arg \max } Q_{n}\left(\tau^{n}, a^{n}\right)\right).
     \end{aligned}
\end{equation}

QMIX extends VDN by factorizing the global value function $Q_{\rm tot}^{\mathrm{QMIX}}(\boldsymbol{\tau}, \boldsymbol{a})$ as a monotonic combination of the agents' local value functions $\left[Q_i(\tau^i, a^i)\right]_{i=1}^n$:
\begin{equation}
     \forall i \in \mathcal{N}, \frac{\partial Q_{\rm tot}^{\mathrm{QMIX}}(\boldsymbol{\tau}, \boldsymbol{a})}{\partial Q_{i}\left(\tau^{i}, a^{i}\right)}>0.
\end{equation}

We mainly implement Fastap on QMIX for its proven performance in various papers. QMIX uses a hyper-net conditioned on the global state to generate the weights and biases of the local Q-values and uses the absolute value operation to keep the weights positive to guarantee monotonicity.

\paragraph{PEARL~\citep{PEARL}:} This baseline comes from single-agent and meta-learning settings. It aims to represent the environments according to some hidden representations. Concretely, PEARL utilizes the transition data as context to infer the feature of the environment, which is modeled by a product of Gaussians. When it is applied to MARL tasks, the PEARL module is adopted and optimized for local context encoders of each individual controllable agent. 

\paragraph{ESCP~\citep{escp}:} As a single-agent reinforcement learning algorithm that aims to recognize and adapt to new environments rapidly when encountering a sudden change in environments, the optimization objective Eqn.~\ref{obj_escp} is applied to optimize a context encoder. To cater to the framework and specific tasks in MARL, the history is not truncated, and each controllable agent is equipped with a local encoder.

\paragraph{LIAM~\citep{LIAM}:}
A method equips each agent with an encoder-decoder structure to predict other agents' observations $\boldsymbol o^{-1}_t$ and actions $\boldsymbol a^{-1}_t$ at current timestep based on its own local observation history $\tau_t=\{o_{0:t}\}$. The encoder and decoder are optimized to minimize the mean square error of observations plus the cross-entropy error of actions. To fit in the MARL setting in our work, local context encoders of controllable agents will be asked to predict the teammates' observations and actions based on their local trajectories. The mean value of their loss is used to optimize the encoders.

\paragraph{ODITS~\citep{ODITS}:}
Unlike the previous two methods that predict the actual behaviors of teammate agents, ODITS improves zero-shot coordination performance in an end-to-end fashion. Two variational encoders are adopted to improve the coordination capability. The global encoder takes in the global state trajectory as input and outputs a  Gaussian distribution. A vector $z$ is sampled and fed into hyper-network that maps the ad hoc agent's local utility $Q_i$ into global utility $Q_{\rm tot}$ to approach the global discounted return. The local encoder has a similar structure and the sampled $e$ is fed into the ad hoc agent's policy network. The encoders are updated by maximizing the return, together with the mutual information of the two context vectors conditioned on the local transition data in an end-to-end manner. As ODITS considers only a single ad hoc agent, we also equip each controllable agent with a local trajectory encoder and maximize the mean of mutual information loss to fit in our MARL's setting.

\subsection{Relevant Environments}
\paragraph{Level-Based Foraging (LBF)~\citep{lbf}:}
LBF is a mixed cooperative-competitive partially observable grid-world game that requires highly coordinated agents to complete the task of collecting the foods. The agents and the foods are assigned with random levels and positions at the beginning of an episode. The action space of each agent consists of the movement in four directions, loading food next to it and a ``no-op'' action, but the foods are immobile during an entire episode. A group of agents can collect the food if the summation of their levels is no less than the level of the food and receive a normalized reward correlated to the level of the food. The main goal of the agents is to maximize the global return by cooperating with each other to collect the foods in a limited time. 

To test the performance of different algorithms in this setting, we consider a scenario with four (at most) agents with different levels and three foods with the minimum levels $l\geq \sum_{i=1}^3 sorted(levels)[i] $ in a $6\times 6$ grid world. Agents have a limited vision with a range of $1$ ($3\times 3$ grids around the agent), and the episode is under a limited horizon of 25. In our Open Dec-POMDP setting, two agents are controllable and will stay in the environment for the whole episode. The number of teammates might be $1$ or $2$, and the policy network will change as well. The rewards that the agents receive are the quotient of the level of the food they collect divided by the summation of all the food levels, as follows: 
    \begin{equation}
    \label{rewardlbf}
      \begin{split}
        r^i = \frac {\rm Food\_with\_Level\_i} { \sum_j \rm Food\_with\_Level\_j}.
        \end{split}
    \end{equation}

\paragraph{Predator-prey (PP)~\citep{maddpg}:} This is a predator-prey environment. Good agents (preys) are faster and receive a negative reward for being hit by adversaries (predators) (-10 for each collision). Predators are slower and are rewarded for hitting good agents (+10 for each collision). Obstacles block the way. By default, there is 1 prey, 3 predators, and 2 obstacles. In our Open Dec-POMDP setting, two predators are controllable and will stay in the environment for the whole episode. The other predator is the uncontrollable teammate whose policy changes suddenly.

\paragraph{Cooperative navigation (CN)~\citep{maddpg}:} In this task, four agents are trained to move to four landmarks while avoiding collisions with each other. All agents receive their velocity, position, and relative position to all other agents and landmarks. The action space of each agent contains five discrete movement actions. Agents are rewarded with the sum of negative minimum distances from each landmark to any agent, and an additional term is added to punish collisions among agents. In our Open Dec-POMDP setting, two agents are controllable and will stay in the environment for the whole episode. The number of teammates might be $1$ or $2$, and the policy network will change as well.

\paragraph{StarCraft II Micromanagement Benchmark (SMAC)~\citep{pymarl}:} SMAC is a combat scenario of StarCraft II unit micromanagement tasks. 
We consider a partial observation setting, where an agent can only see a circular area around it with a radius equal to the sight range, which is set to $9$. We train the ally units with reinforcement learning algorithms to beat enemy units controlled by the built-in AI. At the beginning of each episode, allies and enemies are generated at specific regions on the map. Every agent takes action from the discrete action space at each timestep, including the following actions: no-op, move [direction], attack [enemy id], and stop. Under the control of these actions, agents can move and attack in continuous maps. MARL agents will get a global reward equal to the total damage done to enemy units at each timestep. Killing each enemy unit and winning the combat (killing all the enemies) will bring additional bonuses of $10$ and $200$, respectively. Here we create a map named 10m\_vs\_11m, where 10 allies and 14 enemies are divided into 2 groups separately, and they are spawned at different points to gather together and enforce attacks on the same group of enemies to win this task. Specifically, we control 7 allies to cooperate with 3 other teammates to finish the task, where the number of teammates keeps unchangeable during an episode.
\section{The Architecture, Infrastructure, and Hyperparameters Choices of Fastap}
Since Fastap is built on top of QMIX in the main experiments, we here present detailed descriptions of specific settings in this section, including network architecture, the overall flow, and the selected hyperparameters for different environments.
\subsection{Network Architecture}
In this section, we would give details about the following networks: (1) encoder $E_{\omega_1}$ and decoder $D_{\omega_2}$ in CRP process, (2) trajectory encoder $g_\theta$, $f_{\phi_i}$, and agent networks, and (3) variational distribution $q_\xi$ and teammates modeling decoder $h_{\psi_i}$.

The 8-layer transformer encoder $E_{\omega_1}$ takes global trajectory $\tau=(s_0, \boldsymbol{a}_0,..., s_T)$ as inputs and outputs 16-dimensional behavioral embeddings $v$. The RNN-based decoder $D_{\omega_2}$, consisting of a GRU cell whose hidden dimension is 16, takes $\tau_t^X=(s_0,..., s_t)$ and $v$ as input and reconstructs the action $\boldsymbol{a}_t$.

For the global and local trajectory encoder $g_\theta$ and $f_{\phi_i}$, we design it as a 2-layer MLP and GRU, and the hidden dimension is 64. Then a linear layer transforms the embeddings into mean values and standard deviations of a Gaussian distribution. The context vector will be sampled from the distribution. The global context $z_t$ and state $s_t$ will be concatenated and input into the hypernetwork. As for the local context $e_t^i$, it, together with local trajectory $\tau_t^i$, will be input into the agent $i$'s individual Q network, having a GRU cell with a dimension of 64 to encode historical information and two fully connected layers, to compute the local Q values $Q^i(\tau_t^i, e_t^i, \cdot)$. The local Q values will be fed into the mixing network to calculate TD loss finally.

To maximize the mutual information between local and global context vectors conditioned on the agent $i$'s local trajectory, a variational distribution network $q_\xi$ is used to approximate the conditional distribution. Concretely, $q_\xi$ is a 3-layer MLP with a hidden dimension of 64, and it outputs a Gaussian distribution where the predicted local context vector will be sampled. The agent modeling decoder $h_{\psi_i}$ is divided into two components including $h_{\psi_i}^o$ and $h_{\psi_i}^a$, where each one is a 3-layer MLP. Mean squared loss and maximum likelihood loss are calculated to optimize the objective, respectively.

\subsection{The Overall Flow of Fastap}
To illustrate the overall flow of Fastap, we first show the CRP-based infinite mixture procedure in Alg.~\ref{alg1}. A teammate group can be generated via any MARL algorithm, and we store the small batch of trajectories into a replay buffer $\mathcal{D}_k$ (Line 2\textasciitilde 3). The encoder and decoder are trained to force the learned representation to precisely capture the behavioral information and precisely estimate the predictive likelihood (Line 4). Afterward, the CRP prior and predictive likelihood are calculated to determine the assignment of the newly generated teammate group $m^*$ (Line 5\textasciitilde 7). Then, we update the existing cluster or instantiate a new cluster based on the assignment (Line 8\textasciitilde 17).

The training process of Fastap is also shown in Alg.~\ref{alg2}. During the trajectory sampling stage, we first sample a teammate group from the cluster and fix it in this episode. The teammate group pairs with the controllable agents and they make decisions together (Line 3\textasciitilde12). To train the agent policy networks and the context encoders, the moving average values of context vectors are updated and the optimization objectives are calculated (Line 14\textasciitilde22). Besides, we present the testing process in Alg.~\ref{alg3}, where teammates might change suddenly. A sudden change distribution $\mathcal{U}$ controls the waiting time that determines the changing frequency (Line 5\textasciitilde 12).
\begin{algorithm}[!ht]
    \caption{Fastap: CRP-based infinite mixture procedure}
    \label{alg1}
    \textbf{Input}: concentration param $\alpha$, num of teammate groups generated in one iteration $L$, number of teammate groups generated so far $K$, number of clusters instantiated so far $M$, encoder $E_{\omega_1}$, decoder $D_{\omega_2}$.
    \begin{algorithmic}[1] 
        \FOR{$k=K+1,..,K+L$}
            \STATE Generate the $k^{\text{th}}$ teammate group.
            \STATE Sample small batch of trajectories $\tau_k$ of the $k^{\text{th}}$ teammate group and store them into $\mathcal{D}_k$.
            \STATE Train $E_{\omega_1}$ and $D_{\omega_2}$ according to $\mathcal{L}_{\text{model}}$ in Eqn.~4. 
            \STATE Calculate the CRP prior $P(v_k^{(m)}), m=1,2,...,M+1$ according to Eqn.~2.
            \STATE Calculate the predictive likelihood $P(\tau_k^Y|\tau_k^X;v_k^{(m)}), m=1,2,...,M+1$ according to Eqn.~3.
            \STATE $m^*=\arg\max_{m}P(v_k^{(m)})P(\tau_k^Y|\tau_k^X;v_k^{(m)})$.
            \IF{$m^*\leq M$}
                \STATE Assign the $k^{\text{th}}$ teammate group to the $m^*$ cluster.
                \STATE Update the cluster center $\bar v^{m^*}=\frac{n^{(m^*)}\bar v^{m^*}+v_k}{n^{(m^*)}+1}$.
                \STATE Update the counter of the cluster $m$: $n^{(m^*)}=n^{(m^*)}+1$.
            \ELSE
                \STATE Initialize the $M+1^{\text{th}}$ cluster with the $k^{\text{th}}$ teammate group.
                \STATE Initialize the cluster center $\bar v^{M+1}=v_k$.
                \STATE Initialize the counter of the cluster $M+1$: $n^{(M+1)}=1$.
                \STATE Update $M=M+1$.
            \ENDIF
        \ENDFOR
        \STATE Update $K=K+L$.
    \end{algorithmic}
\end{algorithm}

\begin{algorithm}[!ht]
    \caption{Fastap: training process}
    \label{alg2}
    \textbf{Input}: controllable agent policy networks $\{\pi^i\}_{i=1}^n$, global trajectory encoder $g_\theta$, local trajectory encoders $\{f_{\phi_i}\}_{i=1}^n$, teammate group clusters $\mathcal{C}$, number of clusters instantiated so far $M$, episode length $T$, number of sampled episodes $sample\_num$, environment $env$.
    \begin{algorithmic}[1] 
        \STATE Initialize moving average $\bar z^m=\boldsymbol{0}, m=1,...,M$.
        \STATE Initialize moving average $\bar e^{m,i}=\boldsymbol{0}, m=1,...,M; i=1,..,n$.
        \FOR{$l=1,...,sample\_num$}
            \STATE sample teammate group from $\mathcal{C}$ belonging to the $m^{\text{th}}$ cluser.
            \STATE $s_0^m = env.start()$.
            \FOR{$t=0,...,T$}
                \STATE $e_t^{m,i}=f_{\phi_i}(\tau_t^{m,i}),\quad i=1,...,n$.
                \STATE $a_t^{m,i} = \pi^i(\tau_t^{m,i}, e_t^{m,i}),\quad i=1,...,n$.
                \STATE$\boldsymbol{a}^m_t=(a_t^{m,i})_{i=1}^n$. \texttt{\small // controllable agents decision-making}
                \STATE $\boldsymbol{\bar a}^m_t=\boldsymbol{\bar \pi}^m(\boldsymbol{\bar \tau}_t^m)$. \texttt{\small // uncontrollable teammates decision-making}
                \STATE $s_{t+1}^m, r_t^m=env.step(\langle\boldsymbol{a}^m_t, \boldsymbol{\bar a}^m_t\rangle)$.
            \ENDFOR
            \STATE Add trajectory to the replay buffer $\mathcal{D}$.
            \FOR{$m=1,..,M$}
                \STATE Sample $bs$ trajectories from $\mathcal{D}$.
                \STATE Calculate estimated Q-values and context vectors $z^m_t=g_\theta(\tau_t^m), e_t^{m,i}=f_{\phi_i}(\tau_t^{m, i}),\quad t=0,...,T$.
                \STATE Update $\bar z^m=\eta\text{sg}(\bar z^m)+(1-\eta)\text{mean}(z_t^m)$.
                \STATE Update $\bar e^{m,i}=\eta\text{sg}(\bar e^{m,i})+(1-\eta)\text{mean}(e_t^{m, i})$.
                \STATE Optimize agent Q networks according to $\mathcal{L}_{\text{TD}}$.
            \ENDFOR
            \STATE Optimize $g_{\theta}$ according to $\mathcal{L}_{\text{ADAP}}$ in Eqn.~6.
            \STATE Optimize $\{f_{\phi_i}\}_{i=1}^n$ according to $\mathcal{L}_{\text{DEC}}$ in Eqn.~11.
        \ENDFOR
    \end{algorithmic}
\end{algorithm}

\begin{algorithm}[!ht]
    \caption{Fastap: testing process}
    \label{alg3}
    \textbf{Input}: controllable agent policy networks $\{\pi^i\}_{i=1}^n$, local trajectory encoders $\{f_{\phi_i}\}_{i=1}^n$,  episode length $T$, number of test episodes $test\_num$, environment $env$, sudden change distribution $\mathcal{U}$, teammates set $\mathcal{\bar N}$.
    \begin{algorithmic}[1] 
        \FOR{$l=1,...,test\_num$}
            \STATE Sample teammate policy $\boldsymbol{\bar \pi}$ from $\mathcal{\bar N}$.
            \STATE $s_0 = env.start()$.
            \FOR{$t=0,...,T$}
                \IF{$t=0$}
                    \STATE Sample waiting time $u_0\sim \mathcal{U}$.
                \ELSE
                    \STATE Update waiting time $u_t=u_{t-1}-1$.
                    \IF{$u_t\leq 0$}
                        \STATE Re-sample $u_t\sim \mathcal{U}$.
                        \STATE Re-sample teammate policy $\boldsymbol{\bar \pi}$ from $\mathcal{\bar N}$.
                    \ENDIF
                \ENDIF
                \STATE $e_t^{i}=f_{\phi_i}(\tau_t^{i}),\quad i=1,...,n$.
                \STATE $a_t^{i} = \pi^i(\tau_t^{i}, e_t^{i}),\quad i=1,...,n$.
                \STATE$\boldsymbol{a}_t=(a_t^{i})_{i=1}^n$. \texttt{\small // controllable agents decision-making}
                \STATE $\boldsymbol{\bar a}_t=\boldsymbol{\bar \pi}(\boldsymbol{\bar \tau}_t)$. \texttt{\small // uncontrollable teammates decision-making}
                \STATE $s_{t+1}, r_t, done=env.step(\langle\boldsymbol{a}_t, \boldsymbol{\bar a}_t\rangle)$.
            \ENDFOR
        \ENDFOR
    \end{algorithmic}
\end{algorithm}

Our implementation of Fastap is based on the EPymarl\footnote{\url{https://github.com/oxwhirl/epymarl}}~\citep{lbf} codebase with StarCraft 2.4.6.2.69223 and uses its default hyper-parameter settings (e.g.,   $\gamma=0.99$). The selection of other additional hyperparameters for different environments is listed in Tab.\ref{table_hyper}.
\begin{table*}
    \centering
    \resizebox{\textwidth}{!}{
    \begin{tabular}{|l|cccc|}
    \hline
    \diagbox[width=24em]{Hyperparameter}{\begin{tabular}[c]{@{}l@{}}Environment\\\end{tabular}}               & Level-Based Foraging   & Predator-prey    & Cooperative navigation    & 10m\_vs\_14m  \\ \hline
    concentration hyperparameter $\alpha$           & $0.5$   & $2.5$   & $2.5$   & $0.5$   \\
    number of teammate groups generated in one iteration $L$ & $4$     & $1$     & $1$     & $2$     \\
    radius hyperparameter $\kappa$       & $80$   & $80$    & $80$    & $80$    \\
    moving average hyperparameter $\eta$ & $0.01$  & $0.01$  & $0.01$  & $0.01$  \\
    $\alpha_{\text{GCE}}$                         & $1$     & $0.4$   & $0.4$   & $10$    \\
    $\alpha_{\text{LCE}}$                         & $1$     & $0.4$   & $0.4$   & $10$    \\
    $\alpha_{\text{MI}}$                          & $0.001$ & $0.001$ & $0.001$ & $0.001$ \\
    $\alpha_{\text{REC}}$                         & $0.1$   & $0.2$   & $0.2$   & $0.2$   \\
    dimension of local context vector $e$                       & $4$     & $16$    & $4$     & $8$     \\
    dimension of global context vector $z$                       & $6$     & $20$    & $6$     & $16$    \\ \hline
    \end{tabular}}
    \caption{Hyperparameters in the experiments.}
    \label{table_hyper}
\end{table*}

\bibliography{ref}

\end{document}